\documentclass{emulateapj}
\usepackage{apjfonts,psfig,epsfig,lscape}
\begin{document}

\slugcomment{AJ, in press}

\title{A Comparison of Elemental Abundance Ratios in Globular Clusters, 
Field Stars, and Dwarf Spheroidal Galaxies}

\shortauthors{Pritzl, Venn, \& Irwin}
\shorttitle{Elemental Abundance Comparisons}


\newcommand\teff{T$_{\rm eff}$}
\newcommand\logg{log g }
\newcommand\kms{km\,s$^{-1}$}
\newcommand\afe{[$\alpha$/Fe] \,}

\newcommand\etal{{\rm et al.\,}}
\newcommand\eg{{\it e.g. }}
\newcommand\ie{{\it i.e. }}

\author{Barton J. Pritzl and Kim A. Venn}
\affil{Department of Physics and Astronomy, Macalester College, 1600 
Grand Ave., Saint Paul, MN 55105 
\\email: pritzl@macalester.edu; venn@macalester.edu}
\author{Mike J. Irwin}
\affil{Institute of Astronomy, University of Cambridge, Madingley Road, Cambridge, CB3 0HA, UK 
\\email: mike@cam.ast.ac.uk}

\begin{abstract}
We have compiled a sample of globular clusters with high quality stellar 
abundances from the literature to compare to the chemistries of stars in 
the Galaxy and those in dwarf spheroidal galaxies.   Of the 45 globular clusters 
examined, 29 also have kinematic information.  Most of the globular clusters 
belong to the Galactic halo, however a signficant number have disk kinematics or 
belong to the bulge.  Focusing on the \afe and light r-process element ratios, 
we find that most globular cluster stars mimic those of the field stars of similar 
metallicities, and neither clearly resembles the presently available stellar abundances in 
the dwarf galaxies (including the globular clusters in the Large Magellanic Cloud).  
The exceptions to these general elemental ratio comparisons are already known in the 
literature, e.g., $\omega$~Centauri, Palomar~12, and Terzan~7 associated with the 
Sagittarius remnant, and Ruprecht~106 which has a high radial velocity and low 
\afe ratio.  A few other globular clusters show more marginal peculiarities.  The most notable one 
being the halo cluster M68 which has a high Galactocentric rotational velocity,  a 
slightly younger age, and a unique [Si/Ti] ratio.  The [Si/Ti] ratios decrease with 
increasing [Fe/H] at intermediate metallicities, which is consistent with very massive stars 
playing a larger role in the early chemical evolution of the Galaxy.  The chemical 
similarities between globular clusters and field stars with [Fe/H]$\le-1.0$ suggests a 
shared chemical history in a well mixed early Galaxy.  The differences to the published 
chemistries of stars in the dwarf spheroidal galaxies suggests that neither the 
globular clusters, halo stars, nor thick disk stars had their origins in small 
isolated systems like the present-day Milky Way dwarf satellites.
\end{abstract}


\section{Introduction}

If the Galaxy formed primarily through continuous merging of small dwarf 
systems as demanded by cold dark matter scenarios (e.g., Navarro, Frenk, \& 
White 1997; Klypin \etal 1999; Moore \etal 1999; Bullock, Kravstov, \& Weinberg 
2001; Boily \etal 2004), possibly even until $z=1$, then we ought to be able to trace 
this assembly through the ratio of the elemental abundances in stars.  This is 
because the color-magnitude diagrams of dwarf galaxies in the Local Group have
shown a wide variety of star formation histories and physical properties 
(e.g., Grebel 1997; Mateo 1998; Tolstoy \etal 1998; Dolphin \etal 2003, Skillman 
\etal 2003), which are expected to lead to differences in chemical evolution,
as predicted by many evolution scenarios (e.g., Lanfranchi \& Matteucci 2004; Pagel \& 
Tautvai\u{s}ien\.{e} 1998).  A variety of element ratios are expected in the stars of 
different ages and dwarf galaxy origin.  Therefore, ``chemical-tagging'' (a term from 
Freeman \& Bland-Hawthorn, 2002)  of stellar populations in the Galaxy should be possible 
if significant formation occurred through hierarchical merging.   

The detailed and recent stellar abundances in the smallest, dwarf spheroidal 
galaxies (dSphs; Shetrone \etal 2001, 2003; Tolstoy \etal 2003; Geisler \etal 2005), have shown
their stars tend to have lower \afe abundance ratios than similar metallicity 
Galactic field stars.  The \afe abundance ratio is an important tracer of the relative 
contributions of SNe II/Ia since only SNe II contribute to the production of 
the $\alpha$ elements, whereas both contribute to iron.  The \afe ratio is particularly 
sensitive to this difference at higher metallicities, when SNe Ia have begun to make 
a significant contribution to the chemical evolution of the system;  the Sgr 
dwarf galaxy remnant is an excellent example of this, where the \afe trend is similar 
to that of Galactic halo stars, however offset such that it appears that SNe Ia
have contributed significant amounts of iron at lower metallicities
and \afe is significantly subsolar by [Fe/H]=0 (Bonifacio \etal 2004).  However some 
galaxies, e.g., Draco, show low \afe ratios even at the lowest metallicities 
([Fe/H] $\sim -3.0$; Shetrone, C\^{o}t\'{e}, \& Sargent 2001), which may suggest that 
metal-dependent outflows of SNe II ejecta are also important. 
 
Many metal-poor dSph stars also show low [Y/Eu] abundance ratios, which depends 
on the r- and s-process contributions from SNe II and AGB stars.   Because of 
the slower chemical evolution in the small dwarf galaxies, s-process enrichments 
from {\it metal-poor} AGB stars are important and these stars have lower yields of light
s-process elements, such as Y (Travaglio \etal 2004).   In dSph  stars, the low 
Y is not mimiced by low Ba (as predicted), which results 
in high [Ba/Y] ratios compared to the Galactic field stars (Venn \etal 2004).  
This is most evident around metallicities of [Fe/H] $\sim -2.0$, before the more 
metal-rich AGB stars can also contribute to the s-process yields, even dominate 
the total contribution which can bring the [Ba/Y] ratio, and others, into 
agreement with the Galactic comparison stars (e.g., one metal-rich RGB/AGB star in 
Fornax has lower [Ba/Y] than the metal-poor RGB stars in Fornax, in
excellent agreement with Galactic comparison field stars).

\begin{figure*}[t]
 \centerline{\epsfig{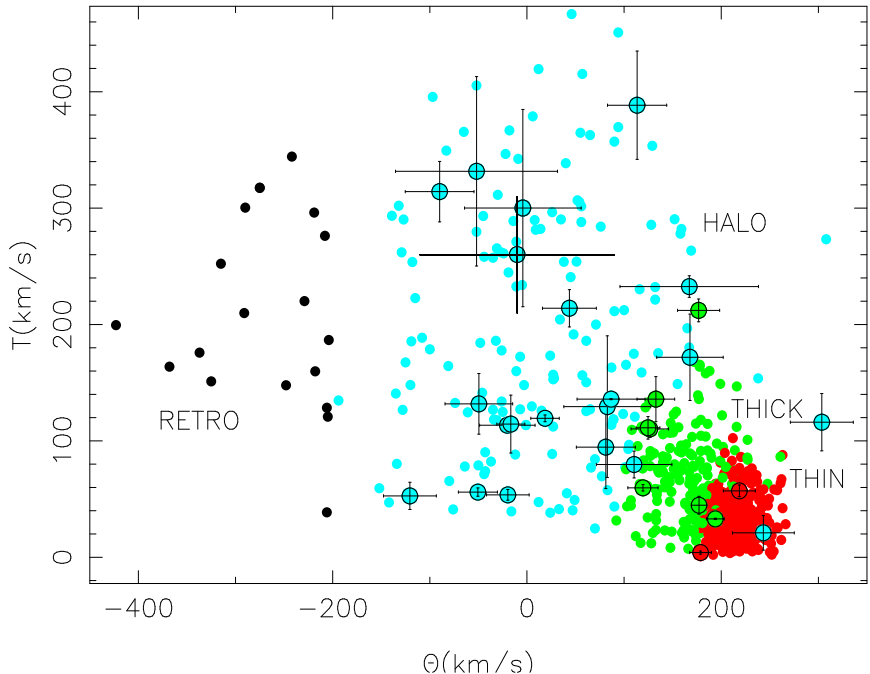}}
 \caption{Toomre diagram showing stellar and globular cluster
populations using kinematic probabilities from velocity ellipsoids;
thin disk (red), thick disk (green), halo (cyan), extreme retrograde
stars (black).  Field stars are points, while globular clusters are
larger black circles filled with a color that refers to its phase space
population.  M54 and Terzan~7 are included in this plot with the
velocity information for the Sagittarius dwarf galaxy remnant.  Ruprecht~106
is included after assuming a range in proper motions (see text).
As expected, a little over half of the globular
clusters have halo kinematics.  See Venn \etal (2004) for field
star data references.}
 \label{Fig01}
\end{figure*}

The chemical comparisons of a large sample of Galactic field stars were compared 
to those in the small dSph galaxies by Venn \etal (2004).   There the stars were 
divided into Galactic components based on their kinematics.   While there is some
marginal evidence that Galactic halo stars with extreme retrograde orbits overlap 
in \afe with the stars in dSphs, this was not evident in the [Ba/Y] ratios.  Thus, it 
appears that no significant component of the Galactic halo, nor the Galactic thick disk 
which was also examined, could have formed from the mergers of these small dwarf galaxies.  
However, these comparisons were based on the available detailed abundances in 
seven dSph galaxies, and primarily from stars in their central fields.  Early results 
from the much broader CaT survey in three dSphs (Sculptor, Fornax, and Sextans from the 
DART survey; e.g., see Tolstoy \etal 2004 for Sculptor) shows that the stars in 
their central fields tend to have higher metallicities than most of the stars out near 
their tidal radii.  It is possible then that these outer dSph field stars have different 
chemical abundance ratios, which may be more similar to the metal-poor 
stars in the Galactic halo; detailed abundance ratios for these outer metal-poor 
dSph stars are pending.  Aside from the selection of stars in the dSphs, 
the comparisons examined by Venn \etal (2004) also did not rule out early merging, 
before the dwarf galaxies had a chance to have a significant and unique chemical 
evolution history; merging with larger dwarf galaxies was also not ruled out.  It is also 
interesting to note that another important physical difference between the large and small 
dSphs galaxies is that the larger dwarf galaxies have higher masses and contain 
globular clusters (GCs; van den Bergh 2000). 

In this paper, we ask how do the GCs fit into this scheme?    Can the
GCs be used as a test of the merging history of large dwarf galaxies in
the formation of our Galaxy?  If some of the Galactic GCs have extragalactic origins, 
then we may assume that they trace the merging history of the {\it large} dwarf 
galaxies, given that GCs are only found in the larger Milky Way dwarf galaxies of 
the Magellanic Clouds, Sagittarius (Sgr), Fornax, and Canis Major (CMa).  We caution 
that the opposite is not true; not all massive dwarf galaxies have GCs.  Therefore we 
can only say that if we find an extragalactic GC that it must have come from a more 
massive dwarf galaxy.  Presumably, differences in star formation histories, chemical evolution, 
and/or the initial conditions (mass, pre-enrichment, or environmental factors) between 
the various dwarf galaxies could 
also leave discernable chemical imprints on their GCs that might be chemically tagged 
when merged into our Galaxy.  It has been shown that the Sgr GCs Terzan~7 (Ter~7) and 
Palomar~12 (Pal~12) follow the abundance trends of the Sgr field stars (e.g., Cohen 2004; 
Tautvai\u{s}ien\'{e} \etal 2004).  In addition, the GCs and field stars analyzed in 
the Large Magellanic Cloud (LMC; Hill \etal 2000; Hill 2004) follow the same abundance trends 
that have lower \afe ratios (O, Ca, and Ti; with the possible exception of Mg) than 
similar metallicity field stars in the Galaxy, which is similar to the field stars 
in the small dwarf galaxies.  Thus, the question is whether there is a population of GCs 
in the Galaxy that have low \afe ratios (and possibly other chemical signatures) that 
could be interpreted as these clusters having formed in dwarf galaxies and later were 
captured through merging.

There have been many papers that present and/or review the abundances 
in GCs, or compare a subset of GCs to one another and/or to Galactic field stars 
(most recently Sneden, Ivans, \& Fulbright 2004; Gratton, Sneden, \& Carretta 2004). 
In general, the Galactic GCs show amazing uniformity in their \afe ratios (plateau 
levels $\sim+0.3$\,dex), and they follow the abundance trends seen in the 
Galactic field stars when plotted as a function of the [Fe/H] values.  This 
seems incredible in terms of understanding hierarchical galaxy formation.  
Carney (1996) showed that there is little or no relationship between \afe 
and age for the GCs.  Given SNe Ia contributions lead to a decrease in the 
\afe ratios with increasing metallicity as seen in the Galactic field stars, 
Carney argued that the lack of a similar turn-down for the GCs implies 
a lack of SNe Ia contributions to the GCs.  Therefore, if the 
timescale of when the Type~Ia supernovae significantly contribute to the 
interstellar medium is short, the ``old'' halo and disk GCs could not share 
a common chemical history and one of the populations must have been later 
accreted.  The exceptions include GCs associated with the Sgr dwarf galaxy, e.g., 
Ter~7 and Pal~12, which have lower \afe ratios than comparison Galactic 
stars (Cohen 2004; Sbordone \etal 2005; Tautvai\u{s}ien\.{e} \etal 2004), and both 
have younger ages than typical Galatic halo GCs.  Also, $\omega$~Centauri 
(Pancino \etal 2002) and Ruprecht~106 (Rup~106; Lin \& Richer 1992; Brown, Wallerstein, \& Zucker 
1997) show peculiar chemical abundances and are thought to be 
captured clusters.  But when it comes to GCs, the more common questions have been 
related to the internal variations in CNO and NaMgAl observed in their red giant stars.  
These variations are attributed to a combination of initial composition differences coupled with
internal mixing mechanisms (Gratton \etal 2004), though others have been exploring 
the possibilities and predictions of enrichments from early AGB stars during the 
cluster formation process (Cottrell \& Da~Costa 1981).  Because these abundance
anomalies are seen only in GCs and never in field stars, including in dwarf galaxy
field stars\footnote{One report of a star in the Sgr dwarf galaxy remnant by Smecker-Hane \& 
McWilliam (2005).  However the metallicity and location of their metal-poor stars make 
them possible members of the M54 GC (Bonifacio \etal 2004).},  Shetrone \etal (1998, 2001, 2003)
concluded that the GCs cannot have formed from the small dwarf galaxies, e.g., during
the merging event.   However, GCs formed in large dwarf galaxies that later
merged into the Galactic halo and survived has not been ruled out.

In this paper, we re-examine the element ratios in as many Galactic GCs as 
available in the literature to compare with the Galactic field stars and 
stars in the dwarf galaxies.  The goal is to indentify candidate GCs that 
show signatures of extragalactic origins either chemically or ideally both 
chemically and kinematically.  While it has been found that smaller dwarf 
galaxies (such as the present-day Milky Way dSph satellites) have 
contributed little or nothing to the Galactic halo (at least from those stars as 
summarized in Venn \etal 2004), the GCs should examine the contributions of larger 
dwarf galaxies.  The field stars with detailed abundance ratios 
include only stars that are currently in the solar neighborhood, while 
the GCs sample all parts of the Galaxy, including the bulge.
Of particular interest are the \afe ratios, as well as the light r-process 
ratios which have been successful in separating the Galactic field stars
from stars in the dSphs.  We study the GC abundance ratios with regards to their 
kinematic populations having adopted the Galactocentric velocities 
($\Pi,\Theta,W$) from Dinescu \etal (1999a,b, 2000, 2001, 2003, 
and private communications for updates) to separate the GCs into standard 
Galactic components (as discussed in \S3).  By noting which component of the 
Galaxy the clusters belong to, we also try to identify clusters that may have 
unusual abundance signatures for their kinematically assigned Galactic component. 

\begin{figure*}
 \centerline{\epsfig{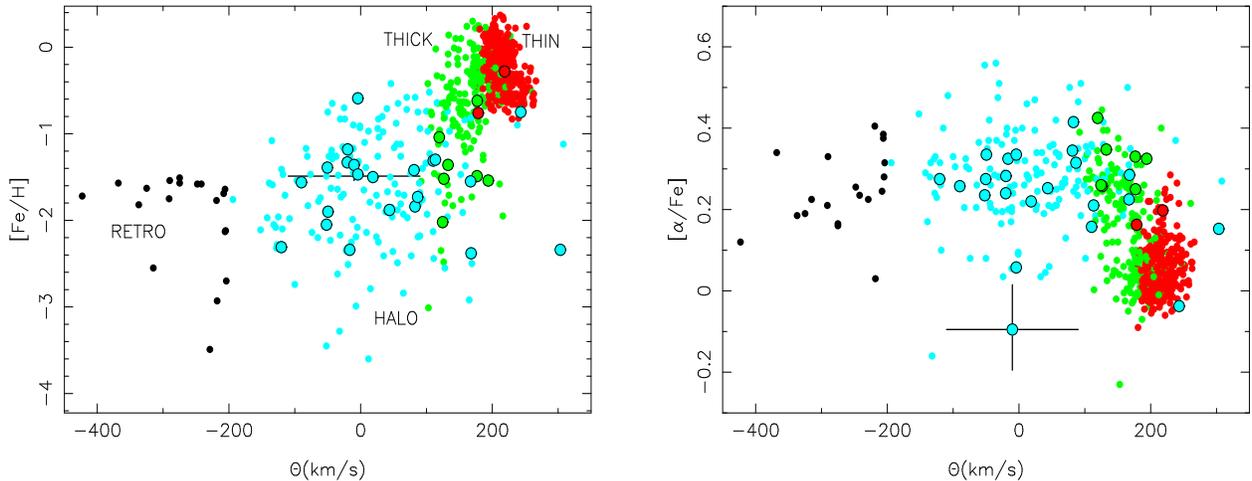}}
 \caption{Variations in [Fe/H] and \afe as a function of the Galactocentric rotational
velocity ($\Theta$) using the same symbols as in Fig.~1.  The field
stars show a large range in the thick disk and halo components, which
also significantly overlap one another.  As in Fig.~1, M54 and Terzan~7
are included in this plot with the velocity information for the
Sagittarius dwarf galaxy remnant, and Ruprecht~106 is included after
assuming a range in proper motions (reflected by the error bar; see text).
The globular clusters tend to have a more narrow range in [Fe/H]
and \afe (where \afe=[(Ca+Ti)/2Fe]), with only a few exceptions
which are discussed in \S3 and \S5.}
 \label{Fig02}
\end{figure*}

\section{The Database} 

Detailed elemental abundances are available for 45 GCs; 32 clusters with $\ge$3 
stars analyzed.  These clusters are listed in Table~1, along with the number 
of stars analysed in each cluster and references.  The literature was 
searched for high-resolution abundances determined within the past 20 years.  
The oldest analysis included here is by Gratton, Quarta, \& Ortolani (1986) for 
three GCs.  A majority of the analyses are from the past five years.  We have 
found that studies from older than 1990 used lower resolution spectra ($\sim15000$).  
More modern studies that use lower resolution (18000 or less) include McWilliam, Geisler, \& Rich 
(1992), Mishenina, Panchuck, \& Samus (2003), Cavallo, Suntzeff, \& Pilachowski (2004).
We consider the abundance ratios coming from these studies as being less reliable.   
In most cases when abundance ratios based on higher resolution spectra are available, 
those from lower resolution spectra are not included in the final mean ratios.

Table~2 lists the elemental abundances and their standard error of
the mean for select elements from Mg to Eu for each of the GCs in Table~1.
We adopted the solar abundances and $\log\,gf$ values used by the Lick-Texas group 
as the standard values (solar abundances, Grevesse \& Sauval 1998; $\log\,gf$, 
see Table~3 in Fulbright 2000 for example).  All abundance ratios were corrected to these 
values (observed - adopted) in an effort to standardize the different 
datasets\footnote{Unless no solar abundances and/or $\log\,gf$ values were given 
in the published article.}.  Details on the adjustments are given in \S4.  
In Table~2, the final individual abundance ratios relative to Fe (shown in bold) were 
averaged together by weighted 
averages according to the number of stars that were used in each study.  The \afe and 
comparison ratios, e.g. [Ba/Y], were calculated using the final weighted-mean 
ratios.  The italicized ratios were not used in calculating the final weighted mean abundance 
ratios.  The GCs in Table~2 span a range of $-2.38$ (M15) $\le$ [Fe/H] $\le -0.06$ 
(NGC 6528).  In a majority of the cases, the [FeI/H] and [FeII/H] ratios were averaged 
when both were available.  Similarly, the [Ti/Fe] ratio is the
average of the [TiI/Fe] and [TiII/Fe] ratios when both were given for the cluster,
otherwise it is the [TiI/Fe] ratio.  Ideally, it would be best to match ionization states 
when determining the Ti ratios, e.g., [TiI/FeI] and [TiII/FeII].  However, in most cases 
not enough information was presented to determine these values.  Many papers 
give the final abundances [TiI/Fe] and [TiII/Fe] with no clear explanation how [FeI/H] 
and [FeII/H] were averaged together to determine [Fe/H].  As a result, we 
used the mean [Fe/H] ratio when determining the Ti ratios.  Our selection of elements 
was based on those determined regularly in GC stars, and those that have been 
useful in the analysis of stars in dSph galaxies.  

Variations in CNO, Na, Mg, and Al are regularly found in clusters from star-to-star, 
and are usually attributed to mixing with CNO-cycled gas (e.g., Kraft 1994; Sneden 
2000; Freeman \& Bland-Hawthorn 2002; Sneden \etal 2004; Gratton \etal 2004).  
We have avoided these elements in this paper (with the exception of magnesium\footnote{Where 
mixing with CNO-cycled gas typically results in  $\Delta$Mg$\le 0.3$~dex (see 
discussion in \S5.1)} given its relevance as a true $\alpha$ element indicator)
because we are primarily concerned with global signatures of galaxy formation and not 
stellar or chemical evolution within GCs.

For abundances in Galactic field stars, we have adopted those in
Table~2 in Venn \etal (2004; see references therein).   The abundances
in seven dSph galaxies are from Shetrone \etal (2001, 2003)
and Geisler \etal (2005).   Stellar abundances for 26 stars in the 
Sgr dwarf galaxy remnant have also been added from Bonifacio \etal 
(2000; 2004) and Smecker-Hane \& McWilliam (2005).

\section{Globular Cluster Kinematic Assignments}

We have determined the Galactic stellar population component for 29 GCs based on a
calculation of their phase space.   Phase space distribution functions
have been determined using the Galactocentric velocity vector components 
(V($\Pi,\Theta,W$) in \kms) from Dinescu \etal (1999ab, 2000, 2001, 2003, and 
private communications for updates) and the Galactocentric positional vector components 
(R(X, Y, Z) in kpc; Harris 1996).  Firstly, we computed the probability for a GC to be 
associated with the Galactic thin disk, thick disk, or halo from its velocity vector using 
a standard Bayesian classification scheme and Galactic Gaussian velocity ellipsoid 
components from Dehnen \& Binney (1998; thin disk), Soubiran, Bienaym\'{e}, 
\& Siebert  (2003; thick disk), and Chiba \& Beers (2000; halo).  This method 
is the same as was used for the Galactic field stars by Venn \etal (2004).  
Secondly, the probability for a GCs to be associated with these Galactic
components was determined from its positional vector.   For this, a
standard Galactic model (Robin, Reyle\'{e}, Derri\`{e}re, \& Picaud 2003) was chosen with a
maximum halo extent of R = 150 kpc, a softening parameter $\rho$=1.0 (which is used 
to stop the density of the halo from going to infinite at the Galactic Center), 
and adopting an asympotitic r$^{-3}$ halo profile.   This was not 
necessary for the Galactic field stars studied by Venn \etal (2004)
since it is a good assumption that those stars are in the solar 
neighbourhood quadrant.  The kinematic and positional probabilities were 
combined for a phase space determination of the final probabilities for each 
GC to be associated with each Galactic component.  Finally, we allocated 
any GCs within R = 2.7 kpc of the Galactic Center to a bulge component unless the 
kinematics placed it in another component, and did 
not probe more deeply into variations between bulge GCs and/or those associated 
with a bar (e.g., Dinescu \etal 2003).  

Table~3 lists the velocity and positional vector components for each
GC, as well as which Galactic component they belong to.   We have also 
added kinematic information for M54 and Ter~7 which are embedded in the Sgr 
dwarf galaxy remnant; thus, on the assumption that these clusters are associated 
with the Sgr remnant, then we have adopted the known kinematics for Sgr (Ibata \etal 1997) 
as a reasonable approximation to their $\Pi,\Theta,W$ velocities.  This allows 
us to highlight these clusters in our abundance ratios and kinematic analyses.  
Rup~106 is another cluster that does not have kinematic information, however it 
has a large galactocentric radial velocity ($-232$ \kms; Harris 1996).   
Adopting a plausible range of proper motions (0",$\pm$1" in RA and DEC), 
we investigate its potential space velocities ($UVW$ and $\Pi,\Theta,W$).   In 
all cases, the Galactocentric radial velocity, $\Pi$, is quite large ($-200$ 
to $-330$ \kms; see Table~4) implying Rup~106 is a member of the Galactic halo, 
possibly on a plunging orbit that would be consistent with a captured cluster.  
The final column in Table~3 lists the GC classifications from (Mackey \& 
Gilmore 2004) which are based on the physical properties of GCs.  While 
overall there is good agreement between the two classifications, there are 
some key differences especially for the clusters which have thick disk 
kinematics, while according to Mackey \& Gilmore they have halo-like properties.

A Toomre diagram for the GCs is shown in Figure~1, where the Galactocentric 
rotational velocity $\Theta$ is plotted against $T$ ($T^2 = \Pi^2 + W^2$).   
To be consistent with the field stars examined by Venn \etal (2004), we plot 
the GCs as colored points according to the assigned Galactic component (e.g., 
cyan = halo, green = thick disk, red = thin disk).  GCs assigned to the bulge 
are not shown in Fig.~1.   Only three clusters stand out in this diagram; 
Pal~12 has thin disk kinematics however its positional vector places it in the 
halo (thus cyan colored GC amongst the thin disk field stars), M68 which has 
an unusually high Galactocentric rotational velocity ($\Theta =+303$ \kms), and M22 is a thick 
disk cluster with an unusually high $T$ component (+212 \kms).  

Figure~2 shows the distribution in [Fe/H] and \afe vs.\ $\Theta$ for the 
GCs compared with the field stars in the solar neighborhood.  Venn \etal (2004) 
commented on the large and overlapping distribution in these abundances for 
the field stars in each of the Galactic components.  For the \afe plot, we 
have averaged the mean abundances of Ca and Ti per GC (note that this differs 
slightly from the field star analysis by Venn \etal 2004 where \afe represented 
an average of Mg, Ca, and Ti).  The most outstanding GCs are Pal~12 and Rup~106, 
with very low \afe ratios.   Of course, Pal~12 is associated with the Sgr 
dwarf galaxy remnant and Rup~106 is thought to have been captured possibly from the 
Magellanic Clouds (Lin \& Richer 1992).  For the thick disk clusters, it is interesting to see that 
they are clustered both toward lower [Fe/H] and higher \afe.  Although they are 
within the range of the thick disk field stars, the thick disk GCs do not show as 
wide of a spread in metallicity.  In any case, we conclude that the \afe ratios 
from the GCs in each kinematic component are in good to excellent agreement with 
those of the field stars in the same Galactic component.

\section{Abundance Ratio Corrections for Globular Clusters}

Most of the data on GC stellar abundances comes from the Lick-Texas group.  
Therefore we have adopted the solar abundances they used (Grevesse \& Sauval 1998), 
along with their $\log\,gf$ values (see Table~3 of Fulbright 2000 for references), 
as the fiducial standards on which all abundances are adjusted.  Differences in 
solar abundances and $\log\,gf$ are in the sense of the value from the reference paper 
minus our adopted values.  A number of clusters were not corrected because 
either their solar abundances and $\log\,gf$ values matched the adopted ones (no 
adjustments were made for differences of less than about 0.04~dex)
or no values were given.  These GCs were noted in Table~1.  In the following we note 
each cluster where adjustments were made to the abundance ratios.  No adjustments 
for differences in hyperfine splitting corrections have been made because such 
abundance ratio adjustments are not as simple as those for solar abundances and $\log\,gf$.

{\bf NGC 104 (47 Tucanae):}  Although the values from Gratton, Quarta, \& 
Ortolani (1986) are not used in the final mean abundances, we adjust them to match 
our adopted $\log\,gf$ values.  There are no solar abundances given in their paper, 
so no adjustments were made.  [Mg/Fe], [Si/Fe], [Ca/Fe], [TiI/Fe], and [Ba/Fe] are 
corrected by $+0.22$, $+0.21$, $+0.20$, $+0.12$, and $-0.86$, respectively, assuming the 
$\log\,gf$ values are the same as in Gratton (1987).
The values from Brown \& Wallerstein (1992) are not used in the final mean abundances, 
but we adjusted them to match our adopted solar abundances and $\log\,gf$ values.  
Since the $\log\,\epsilon$ values are given, the ratios are determined directly from those 
values using our adopted solar abundances.  From the $\log\,gf$ values, [Mg/Fe], [Si/Fe], 
and [TiII/Fe] have been corrected by $+0.09$, $+0.05$, and $+0.25$, respectively.
To avaoid confusion, in Table~3 we have combined the results from Gratton \etal (2001) and 
James \etal (2004a) since they are both part of a collaboration that studies the same stars, 
but different element ratios.  The abundances from James \etal are weighted means 
of the turnoff and subgiant star ratios.

{\bf NGC~288, NGC~362, NGC~5897, NGC~6352, and NGC~6362:}  No solar abundance values were given 
in Gratton (1987), so the only adjustments that were made were due to the $\log\,gf$ values.  
[Mg/Fe], [Si/Fe], [Ca/Fe], [TiI/Fe], and [Ba/Fe] are corrected by $+0.22$, $+0.21$, $+0.20$, 
$+0.12$, and $-0.86$.  For NGC~288 and NGC~362, the results from Gratton (1987) are superceeded 
by those in Shetrone \& Keane (2000).

{\bf NGC 2298:}  No changes are needed due to the solar abundance values in McWilliam, Geisler, 
\& Rich (1992).  Only [La/Fe] is adjusted by $+0.09$ due to the $\log\,gf$ values.

{\bf NGC 3201:}  We do not use the Gratton \& Ortolani (1989) ratios because two of the three stars 
are redone by Gonzalez \& Wallerstein (1998).
Examining the solar abundances, adjustments in [Mg/Fe], [TiI/Fe], and [Ba/Fe] 
needed to be made for the Gonzalez \& Wallerstein abundance ratios.  
Further corrections need to be made due to differences in the $\log\,gf$ values for 
[Mg/Fe], [TiI/Fe], and [Eu/Fe].  It happens that the adjustments for the 
differences in the solar abundances and $\log\,gf$ cancel each other out for 
[Mg/Fe] and [TiI/Fe] so that no adjustments are necessary.  The total changes for 
[Ba/Fe] and [Eu/Fe] are $+0.23$ and $+0.17$, respectively.
In determing the Gonzalez \& Wallerstein ratios, we have chosen to use the high 
resolution abundances from 1991 and 1994 for those stars observed multiple times.

{\bf NGC 4590 (M68):}  The results from Shetrone \etal (2003) and Lee, Carney, \& 
Habgood (2004) superceed those from Gratton \& Ortolani (1989).  The abundance ratios 
derived from photometric values in Lee, Carney, \& Habgood were used.

{\bf NGC 5273 (M3):}  The results of Kraft \etal (1993, 1995) were redone by Sneden \etal (2004).  
In addition, the three stars in Shetrone, C\^{o}t\'{e}, \& Sargent (2001) are also done in 
either Sneden \etal or Cohen \& Melendez (2005a).  Therefore, we did not use their abundance 
ratios in the final weighted-mean ratios.  
For the Cohen \& Melendez study, the [Fe/H] values were adjusted 
by $-0.07$ due to a difference in the solar abundance value.  The abundance 
ratios [Mg/Fe], [Ca/Fe], and [Eu/Fe] were adjusted due to differences 
in the $\log\,gf$ values by $-0.10$, $+0.16$, and $-0.08$.
Therefore the final abundance ratios for M3 are a weighted mean of those from Sneden 
\etal (2004) and Cohen \& Melendez (2005a).

{\bf NGC 5466:}  There are no changes to the NGC 5466 abundance ratios.  We note that 
the only available chemical abundances come from the anomalous Cepheid in this cluster.  
This type of variable star can derive from either binary mass transfer or younger stars 
(Demarque \& Hirshfeld 1975; Norris \& Zinn 1975; Renzini, Mengel, \& Sweigart 1977).  
Because of the star's variability, we caution about the reliability of the abundance 
ratios from this single star even though they are consistent with other stars 
and clusters of similar metallicity.

{\bf NGC 5904 (M5):}  Although the values from Gratton, Quarta, \& Ortolani (1986) 
are not used in the final mean abundances, we adjust them to match our adopted 
$\log\,gf$ values.  There are no solar abundances given in the paper, so no 
adjustments can be made.  [Mg/Fe], [Si/Fe], [Ca/Fe], [TiI/Fe], and [Ba/Fe] are corrected 
by $+0.22$, $+0.21$, $+0.20$, $+0.12$, and $-0.86$, respectively, assuming the $\log\,gf$ 
values are the same as in Gratton (1987).
The ratios from Sneden \etal (1992) are not used because the stars were reanalyzed 
by Ivans \etal (2001).  The asymptotic giant branch stars in the Ivans \etal study 
were not included in the mean abundance ratios.
For the Ram\'{i}rez \& Cohen (2003) abundance ratios, adjustments are necessary 
for both the solar abundances and the $\log\,gf$ values.  For the solar abundances, 
[Mg/Fe], [Ca/Fe], [TiI/Fe], [FeI/H], [FeII/H], and [La/Fe] need to be adjusted 
by $-0.12$, $-0.22$, $-0.07$, $-0.08$, $-0.05$, and $-0.08$, respectively.  For 
the $\log\,gf$ values, [Mg/Fe], [Ca/Fe], and [Eu/Fe] need corrections of $-0.08$, 
$+0.16$, and $-0.08$, respectively.  This leads to a total adjustments of 
$-0.20$, $-0.06$, $-0.07$, $-0.08$, $-0.05$, $-0.08$, and $-0.08$ for [Mg/Fe], 
[Ca/Fe], [TiI/Fe], [FeI/H], [FeII/H], [La/Fe], and [Eu/Fe].
After all of the correction were made, the Ivans \etal and Ram\'{i}rez \& Cohen 
ratios were averaged together by a weighted mean.

{\bf NGC~6093 (M80):}  The solar abundances in Cavallo, Suntzeff, \& Pilachowski (2004) matched our adopted 
values.  For the $\log\,gf$ values, correction were made to [Ca/Fe], [TiII/Fe], and [Eu/Fe] 
by $+0.13$, $-0.06$, and $+0.22$, respectively.

{\bf NGC 6121 (M4):}  Although the values from Gratton, Quarta, \& Ortolani (1986) 
are not used in the final mean abundance ratios, we adjust them to match our adopted 
$\log\,gf$ values.  There are no solar abundances given in the paper, so no 
adjustments can be made.  [Mg/Fe], [Si/Fe], [Ca/Fe], [TiI/Fe], and [Ba/Fe] are corrected by 
$+0.22$, $+0.21$, $+0.20$, $+0.12$, and $-0.86$, respectively, assuming the $\log\,gf$ values 
are the same as in Gratton (1987).
The values from Brown \& Wallerstein (1992) are not used in the final mean abundance ratios, 
but we adjusted them to match our adopted solar abundances and $\log\,gf$ values.  
Since the $\log\,\epsilon$ values are given, the ratios were determined directly from those 
values using our adopted solar abundances.  From the $\log\,gf$ values, [Mg/Fe] and [Si/Fe] 
have been corrected by $+0.09$ and $+0.05$, respectively.
Ivans \etal (1999) re-observed one star in Gratton, Quarta, \& Ortolani (1986) and all three stars 
in Brown \& Wallerstein (1992) and therefore supersedes these two earlier studies.

{\bf NGC 6205 (M13):}  The eighteen stars in Kraft \etal (1997) were reanalyzed by Sneden \etal 
(2004) (the latter adopted here).  As such, the [Mg/Fe] values from Sneden \etal supercede 
those from Kraft \etal, while the remaining abundances were used to calculated the final 
weighted-mean ratios.
For Cohen \& Melendez (2005a), the [Fe/H] ratios needed to be corrected by $-0.07$.  
The ratios [Mg/Fe], [Ca/Fe], and [Eu/Fe] need to be corrected due to differences in 
$\log\,gf$ values by $-0.10$, $+0.16$, and $-0.08$, respectively.  
There are four stars that match between the Sneden \etal and Cohen \& Melendez.  
Given this is a small number compared to the total number of stars analyzed, 
we calculated the final abundance ratios from a weighted average of the two studies.

{\bf NGC~6287, NGC~6293, and NGC~6541:}  No adjustments are needed for the Lee \& Carney 
(2002) abundance ratios due to the solar abundances.  However, [Mg/Fe] has been corrected 
by $-0.10$ due to $\log\,gf$ differences.  The ratios derived from photometric values 
were used.

{\bf NGC 6341 (M92):}  Four of the six stars in Shetrone (1996) have been reobserved 
by either Sneden, Pilachoski, \& Kraft (2000) or Shetrone, C\^{o}t\'{e}, \& Sargent (2001).  
Therefore, we will not use the Shetrone (1996) results in calculating the final abundance ratios.  
Although, three of the four stars in Shetrone, C\^{o}t\'{e}, \& Sargent were also observed 
by Sneden, Pilachoski, \& Kraft, we use the ratios for [Mg/Fe], [Y/Fe], and [Eu/Fe] from the 
former study in the final weighted-mean ratios because they were not calculated by the later 
study.

{\bf NGC 6342:}  The spectra in Origlia, Valenti, \& Rich (2005) were taking in the infrared, which 
does not allow for a direct comparison with the adopted Lick-Texas $\log\,gf$ 
values.  No changes are necessary due to the solar abundances.

{\bf NGC 6528:}  No adjustments were made to the Carretta \etal (2001) results due to the 
solar abundances.  [Mg/Fe] was corrected by $-0.18$ because of the $\log\,gf$ 
values.
No adjustments are needed for the Zoccali \etal (2004) ratios due to the 
solar abundances.  Due to differences in the $\log\,gf$ values, [Mg/Fe] and 
[Ca/Fe] need to be adjusted by $-0.13$ and $+0.15$.
The spectra in Origlia, Valenti, \& Rich (2005) were taking in the infrared, which
does not allow for a direct comparison with the adopted Lick-Texas $\log\,gf$
values.  No changes are necessary due to their solar abundances.

{\bf NGC 6553:}  No solar abundances were given in Barbuy \etal (1999), so only adjustments 
were made due to the $\log\,gf$ values.  Their abundance ratios for [Mg/Fe] and [La/Fe] have been 
corrected by $+0.23$ and $-0.09$, respectively.
The only adjustment in the Cohen \etal (1999) ratios is for [Ca/Fe] by $-0.20$ due 
to difference between the adopted solar abundance.  
As a side note, Origlia, Rich, \& Castro (2002) using infrared spectra give a general 
\afe ratio for NGC 6553 of $+0.30$.  We do not include their results in Table~2 because they do not 
give the individual star values.

{\bf NGC 6656 (M22):}  The values from Brown \& Wallerstein (1992) have been adjusted to match 
our adopted solar abundances and $\log\,gf$ values.  [FeI/H] and
[FeII/H] are corrected by $+0.19$ and $+0.13$, respectively.  There is no
change to [Y/Fe] and [La/Fe].  For the other ratios, [Mg/Fe], [Si/Fe], [Ca/Fe], [TiI/Fe], [TiII/Fe],
[Ba/Fe], and [Eu/Fe] are adjusted by $+0.19$, $+0.21$, $-0.05$, $+0.09$, $-0.33$, $+0.34$, and
$+0.15$, respectively.
There is only one star common between both studies.  The final abundance ratios are a 
weighted mean of the two studies.

\begin{figure*}
 \centerline{\epsfig{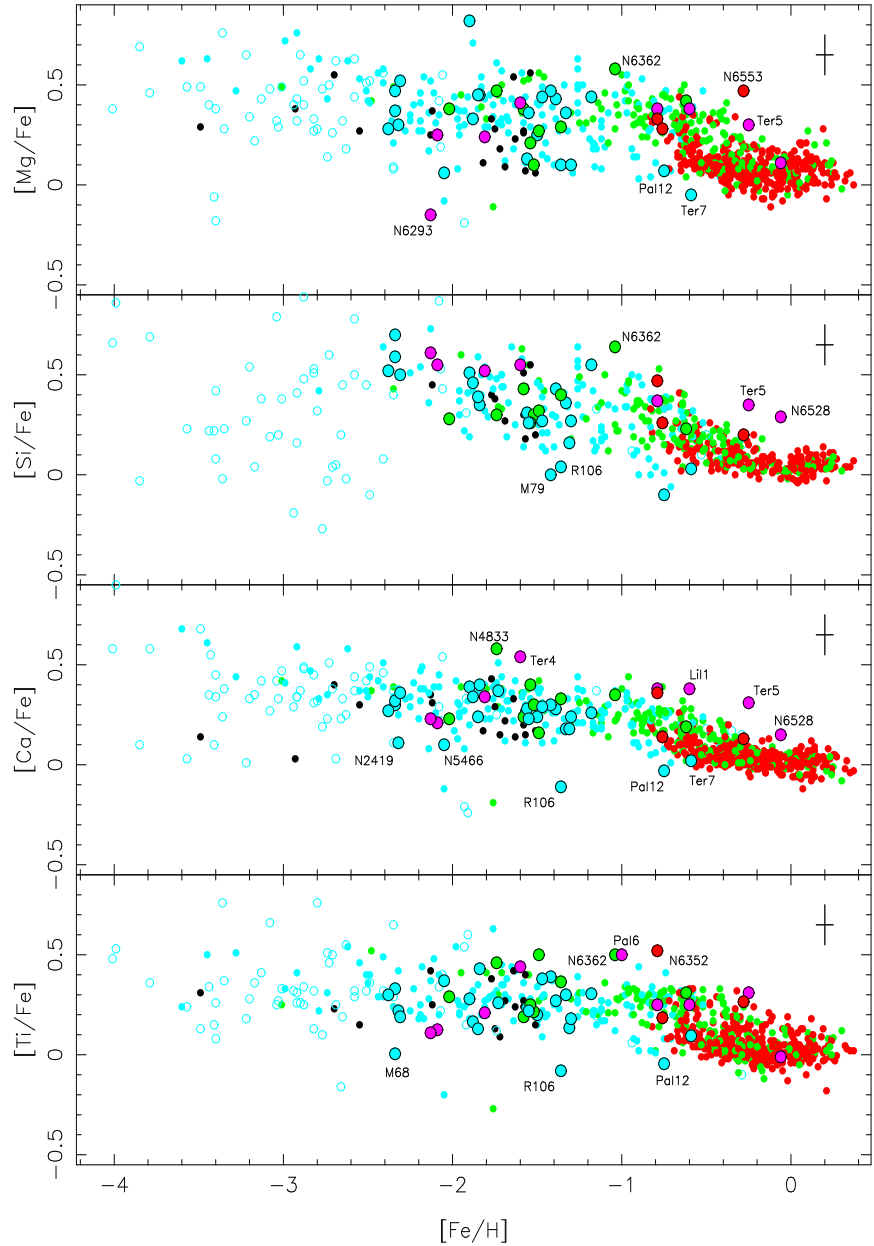}}
 \caption{Abundance ratios for the $\alpha$ elements Mg, Si, Ca, Ti,
in the globular clusters and Galactic field stars (from Venn \etal 2004);
symbols and colors as in Fig.~1.  We have added a bulge component for
four interior clusters (according to Dinescu \etal 2003);   the globular
clusters associated with the bulge are shown in magenta.  Small open
circles are field stars without kinematic information.  The range in
[Mg/Fe] is similar between the field stars and globular clusters, however
several globular clusters stand out in the [Si/Fe], [Ca/Fe] and possibly [Ti/Fe]
plots.  The error bars represent a mean uncertainty of $\pm0.10$ for the
$\alpha$-elements and $\pm0.05$ for [Fe/H].}
 \label{Fig03}
\end{figure*}

{\bf NGC 6715 (M54):}  Due to differences in both the solar abundances and the $\log\,gf$ values, we 
corrected [Mg/Fe], [Si/Fe], [Ca/Fe], [TiI/Fe], [Ba/Fe], [La/Fe], and [Eu/Fe] by 
$+0.29$, $+0.16$, $+0.08$, $+0.09$, $+0.35$, $-0.06$, and $+0.15$, respectively, for 
the Brown, Wallerstein, \& Gonzalez (1999) abundance ratios.  The [FeI/H] and 
[FeII/H] ratios need to be corrected by $+0.10$ and $+0.04$.

{\bf NGC 6752:}  Although the values from Gratton, Quarta, \& Ortolani (1986) are 
not used in the final mean abundances, we adjust them to match our adopted $\log\,gf$ 
values.  There are no solar abundances given in the paper, so no adjustments can be made.  
[Mg/Fe], [Si/Fe], [Ca/Fe], [TiI/Fe], and [Ba/Fe] are corrected by $+0.22$, $+0.21$, $+0.20$, 
$+0.12$, and $-0.86$, respectively, assuming the $\log\,gf$ values are the same as in Gratton (1987).
The only adjustments to the Yong \etal (2003) ratios is for [Mg/Fe] by $-0.11$ due 
to $\log\,gf$ differences.
No changes need to be made to the Cavallo, Suntzeff, \& Pilashowski (2004) ratios for 
the solar abundances.  However, corrections need to be made due to the $\log\,gf$ values.  
[Ca/Fe], [TiII/Fe], and [Eu/Fe] are corrected by $+0.13$, $-0.06$, and $+0.22$.
To avoid confusion, in Table~3 we have combined the results from Gratton \etal (2001) 
and James \etal (2004a) since they are both part of a collaboration that studies the 
same stars, but different element ratios.  The abundances from James \etal are weighted 
means of the turnoff and subgiant star ratios.

{\bf NGC 6838 (M71):}  Although the values from Gratton, Quarta, \& Ortolani (1986) 
are not used in the final mean abundances, we adjust them to match our adopted $\log\,gf$ 
values.  There are no solar abundances given in the paper, so no adjustments can be made.  
[Mg/Fe], [Si/Fe], [Ca/Fe], [TiI/Fe], and [Ba/Fe] are corrected by $+0.22$, $+0.21$, $+0.20$, 
$+0.12$, and $-0.86$, respectively, assuming the $\log\,gf$ values are the same as in Gratton (1987).
There are difference in both the solar abundances and the $\log\,gf$ values 
between our adopted ones and those from Ram\'{i}rez \& Cohen (2002).  [Fe/H] 
has been corrected by -0.06.  The final abundance ratios [Mg/Fe], [Ca/Fe], 
[Y/Fe], and [La/Fe] have been corrected by $-0.07$, $-0.29$, $+0.23$, and 
$-0.14$.
In Table~2, we include the results from Lee, Carney, \& Balachandran (2004), 
which is an infrared study.  Both stars in their study were already observed 
by Sneden \etal (1994) with similar results.  Also, one of the two 
stars was observed in Ram\'{i}rez \& Cohen.  To allow for a better comparison 
between the different datasets, we do not include the ratios from the infrared study 
of Lee, Carney, \& Balachandran in the final weighted-mean ratios.

{\bf NGC 7078 (M15):}  The three stars in Sneden \etal (2000) were also observed by 
Sneden \etal (1997) and Sneden, Pilachowski, \& Kraft (2000).  Still, we include the 
results from Sneden \etal (2000) in the final weighted-mean ratios because they are 
updates from the other two studies.  In addition, ten stars 
are common between the Sneden \etal (1997) and Sneden, Pilachowski, \& Kraft (2000) studies.  
Therefore, in the cases where these two studies have common element ratios, those from 
Sneden, Pilachowski, \& Kraft supercede those from Sneden \etal (1997).

{\bf NGC 7492:}  The [Fe/H] value from Cohen \& Melendez (2005b) needs to be adjusted by 
$-0.05$ due to differences in the solar abundances.  Differences in 
solar abundances also require a change for the [La/Fe] ratio by $-0.08$.  
From the $\log\,gf$ difference, [Mg/Fe], [Ca/Fe], [Eu/Fe], have been 
corrected by $-0.10$, $+0.16$, and $-0.08$, respectively.

\begin{figure*}
 \centerline{\epsfig{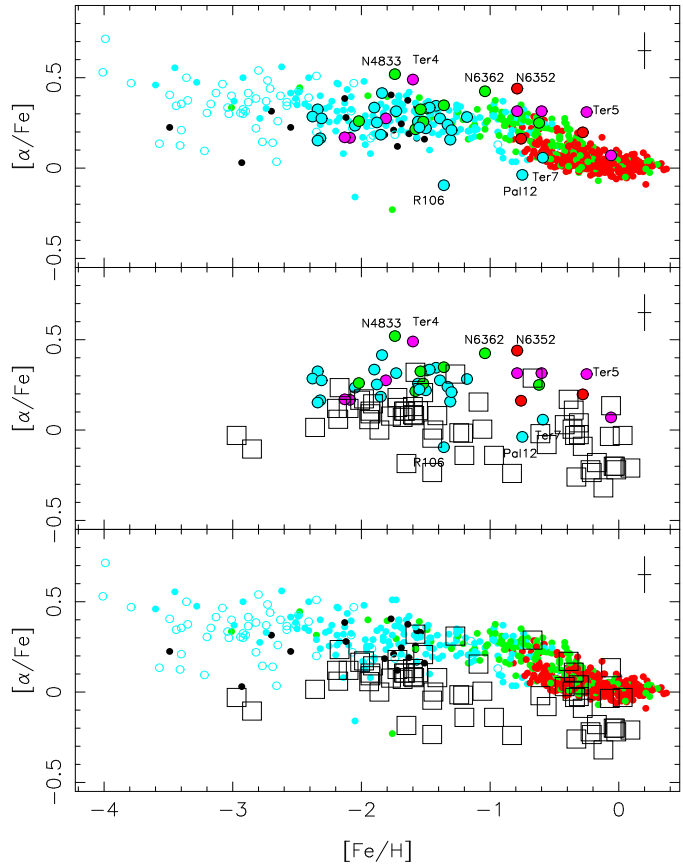}}
 \caption{The \afe abundance ratios (mean of Ca and Ti) for the
globular clusters, as well as the stars in dwarf spheroidals (hollow
squares) and Galactic field stars (from Venn \etal 2002, also Bonifacio
\etal 2004 for Sgr stars); symbols and colors the same as in Fig.~1.  Most
globular clusters show similar \afe ratios as the Galactic field stars.
The exceptions (Palomar~12, Ruprecht~106) show lower \afe ratios
similar to the dwarf spheroidal field stars.  The error bars represent a
mean uncertainty of $\pm0.10$ for the $\alpha$-elements and $\pm0.05$ for [Fe/H].}
\end{figure*}

{\bf Liller 1:}  Although individual stellar abundances were not given in Origlia, Rich, \& 
Castro (2002) for Liller 1, a mean \afe ratio of $+0.03$ was given, with 
[Fe/H]$=-0.3\pm0.2$.  This study was done with infrared spectra and therefore it is not 
possible to compare their $\log\,gf$ values to those from the Lick-Texas group.  The solar 
abundances match our adopted values.

{\bf Palomar 6:}  This cluster was observed by Lee, Carney, \& Balachandran (2004) in the infrared 
and therefore it is not possible to compare their $\log\,gf$ values to those from the Lick-Texas group.  
The solar abundances match our adopted values.  Although three stars were observed, only one star had a [Ti/Fe] value, which is 
$+0.5$ and [FeI/H]$=-1.0\pm0.1$.  No other elements we used in this paper were given.

{\bf Palomar 12:}  The Brown, Wallerstein, \& Zucker (1997) [Mg/Fe] ratio has been corrected by 
$+0.08$ due to $\log\,gf$ differences.
For the Cohen (2004) ratios, [Mg/Fe], [Ca/Fe], [La/Fe], and [Eu/Fe] needed to be 
corrected by $-0.10$, $+0.16$, $-0.09$, \& $-0.08$, respectively, due to differences 
in $\log\,gf$ and the solar abundances.
The two stars in Brown, Wallerstein, \& Zucker were reobserved by Cohen 
so we use only the Cohen results for the final abundance ratios.

{\bf Ruprecht 106:}  The Brown, Wallerstein, \& Zucker (1997) [Mg/Fe] ratio has been corrected by
$+0.08$ due to $\log\,gf$ differences.

{\bf Terzan~4 and Terzan~5:}  The Origlia \& Rich (2004) ratios for Ter~4 and Ter~5 are in the 
infrared and cannot be directly compared to the $\log\,gf$ values from the Lick-Texas group.  
The solar abundances match our adopted values.

{\bf Terzan 7:}  For the Sbordone \etal (2005) ratios, the only correction that needs to be made 
is to the [Mg/Fe] ratio by $-0.08$ due to differences in the $\log\,gf$ values.

\begin{figure*}
 \centerline{\epsfig{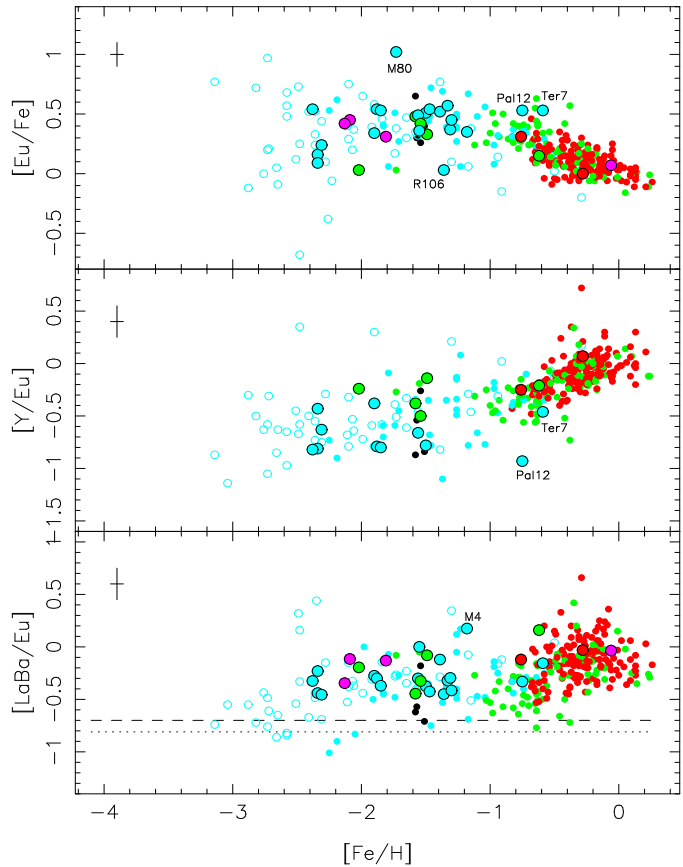}}
 \caption{Abundance ratios for the neutron capture elements, [Eu/Fe] and
the others relative to Eu to examine the r- and s-process contributions.
Again, the globular clusters show similar ratios to the Galactic field
stars (from Venn \etal 2004).   Only Palomar~12 is significantly offset
with a low [Y/Eu] ratio.  The pure r-process estimates from solar system
abundances shown in the bottom panel are from Arlandi \etal (1999; dashed line)
and Burris \etal (2000; dottel line).  We chose to use the estimates for [Ba/Eu]
since even though we average the Ba and La abundance ratios, a majority of values
come from Ba.  The error bars represent a mean uncertainty of $\pm0.10$ for [Eu/Fe],
$\pm0.15$ for the ratios relative to Eu, and $\pm0.05$ for [Fe/H].}
 \label{Fig05}
\end{figure*}

\section{Abundance Ratios}

\subsection{The $\alpha$ elements}

In Figure~3, we plot the [Mg/Fe], [Si/Fe], [Ca/Fe], and [Ti/Fe] ratios for the GCs 
in our sample, as well as the field stars from Venn \etal (2004).  
Sneden \etal (2004) showed that [Ca/Fe] vs.\ [Fe/H] in the GCs 
closely follow the Galactic field stars.  We confirm this, and note 
that this trend holds for Mg, Si, and Ti as well, though with more scatter. 
The range in the [Mg/Fe] ratios is significantly larger
than the other ratios for both the GC and field stars.   The cause
of this dispersion may differ between the GC and field stars though.
In the GCs, the [Mg/Fe] variation may be related to stellar evolution
effects and internal mixing effects, e.g., the typical star-to-star
variation in a GC is 0.3~dex (as determined from the maximum range 
in [Mg/Fe] per GC in our sample).  The dispersion in the field stars has 
been shown not to be related to atmospheric parameters or oscillator strengths, 
but may be due to departures from LTE (Carretta, Gratton, \& Sneden 2000).  

The agreement between these $\alpha$-element ratios in the GCs and field stars is 
best for metal-poor clusters, [Fe/H] $\le -2$.  This implies a uniformity 
in the SNe II yields of $\alpha$-elements and iron, and no contributions 
from SNe Ia (nor AGB stars) as expected.  NGC~2419 (a possible Sgr dwarf 
galaxy GC; Newberg \etal 2003) has a marginally lower [Ca/Fe] ratio, 
though [Mg/Fe], [Si/Fe], and [Ti/Fe] resemble the halo field stars well.  The bulge 
cluster NGC~6293 has signficantly lower [Mg/Fe] and marginally lower [Ti/Fe].  
The halo cluster M68 has lower [Ti/Fe] and higher [Si/Fe] ratios as noted by Lee, Carney, \& 
Habgood (2004), although its [Mg/Fe] and [Ca/Fe] ratios are consistent with 
field stars of similar metallicity.

\begin{figure*}
 \centerline{\epsfig{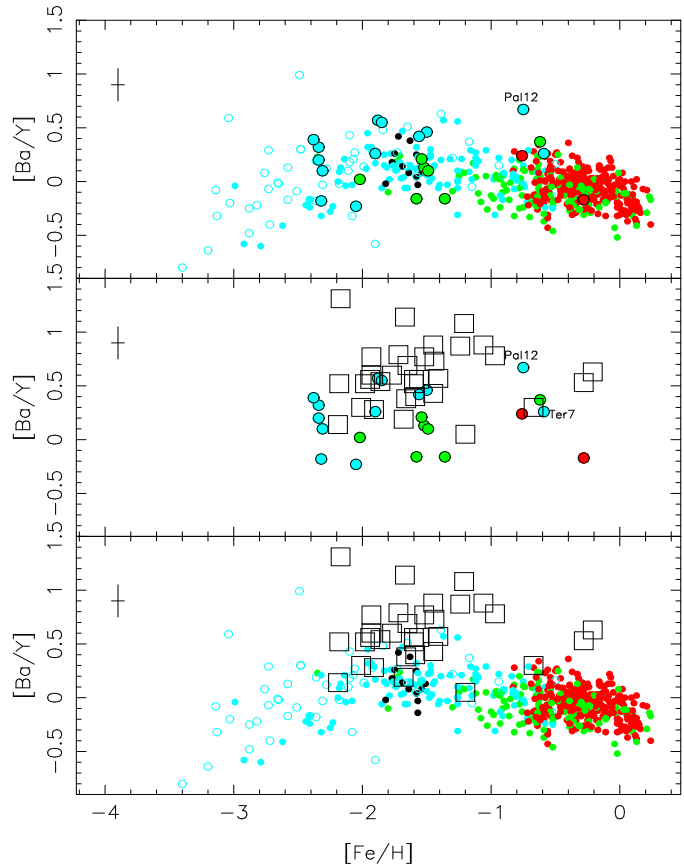}}
 \caption{The [Ba/Y] ratios for the globular clusters, as well as the
stars in dwarf spheroidals and Galactic field stars (from Venn \etal 2004);
same symbols as in Figs.~1 and 4.  This ratio is carefully examined
because of the offset between the stars in dwarf spheroidals and
Galactic field stars, interpreted as metallicity dependent AGB yields
(s-process contribution) and possibly a new or metallicity-dependent
r-process contribution (Travaglio \etal 2004; Venn \etal 2004).
The offset is still visible between the globular clusters and dwarf
spheroidals (central panel), though it is less pronounced for the halo
globular clusters and more clear with the thick disk (green) clusters.
The error bars represent a mean uncertainty of $\pm0.15$ for [Ba/Y]
and $\pm0.05$ for [Fe/H].}
 \label{Fig06}
\end{figure*}

At intermediate metallicities ($-2 \le$ [Fe/H] $\le -1$), there 
is a larger dispersion in the \afe ratios.  The [Mg/Fe] ratios vary by 
0.5~dex, however we do {\it not} suggest that this dispersion is due to 
mixing of GCs from a variety of different dwarf galaxy mergers.  The [Ca/Fe] and 
[Ti/Fe] ratios in the GCs show smaller dispersions (considering Rup~106 as an 
outlier), while the [Si/Fe] ratios have dispersions somewhere in-between.  
This suggests that most of the GCs are in good 
agreement with the field star distribution.  The most outstanding outlier 
is Rup~106 at ${\rm [Fe/H]}=-1.4$.  This GC has been noted for its young age 
(Buonanno \etal 1990, 1993) and Lin \& Richer (1992) speculated that it may 
be a recent capture from the Magellanic Clouds.  Here we see it is also outstanding
in its low ratios for three of the four $\alpha$-elements, typically sub-solar. 
Both NGC~4833 and Ter~4 also stand out with high [Ca/Fe] ratios;
however, the data for NGC~4833 is from only two stars analysed by
Gratton \& Ortolani (1989, which is one of the oldest publications
included in our database), and the data for the highly reddened cluster
Ter~4 are from infrared echelle spectroscopy which requires a slightly
different analysis method and spectral lines from the more commonly used 
optical spectroscopy and optical lines.  NGC~6362 has high [Si/Fe] and [Ti/Fe] ratios, 
but they derive from specta of two stars in Gartton (1987) which is an older 
study and should be reobserved before firm conclusions can be drawn.

In Figure~4 we compare the \afe ratios of the GCs, the Galactic field stars, 
and the stars in dSph galaxies.  The $\alpha$-index is an average of Ca and 
Ti, although only Ca is available for stars in the Sgr dwarf galaxy from 
Bonifacio \etal (2004).  In Venn \etal (2004), Mg, Ca, and Ti were averaged for 
the $\alpha$-index for the Galactic field stars, however here (bottom panel) 
we have not included Mg for field star comparisons since this element can vary significantly 
from star to star in GCs (see \S2).  The mean \afe in most of the GCs
mimics that in the Galactic field stars (top panel), as was seen for the 
individual \afe element abundances in Fig.~3.  Most of the exceptions to 
this good agreement have low [$\alpha$/Fe], in better agreement with the dSph 
galaxies (middle panel).  We also note that amongst the metal-rich Sgr dwarf 
remnant stars, the half with disk-like \afe ratios are from Smecker-Hane \& McWilliam
(2005) whereas the half with sub-solar \afe are from Bonifacio \etal
(2000; 2004).

About half (4 out of 10) of the metal-rich GCs ([Fe/H] $\ge -1.0$) have \afe 
ratios that are not in good agreement with the Galactic field stars. 
Pal~12 has significantly lower \afe ratios, while Ter~5 and NGC~6352 
are significantly higher.  Ter~5 has higher [Mg/Fe], [Si/Fe], and [Ca/Fe] ratios than
most of the metal-rich GCs and field stars, however these abundances
are from  infrared echelle spectroscopy (like Ter~4, see above) and may
suffer from unrecognized systematic offsets when compared with optical
spectroscopic results.  The low \afe ratio for Pal~12 have been discussed 
(Cohen 2004); combined with its kinematics (Dinescu \etal 2000), this provides 
compelling evidence that this cluster was formed in the Sgr dwarf galaxy before 
merging with the Galaxy.  NGC~6352 has a high [Ti/Fe] ratio, but the data come from
an older study (Gratton 1987) and need updating.

\subsection{Neutron Capture Element Ratios}

Just as the \afe ratios are related to variations in the SNe II/SNe Ia 
contributions, the neutron capture element ratios are sensitive to the variations 
in the massive star IMF, variations in SNe II/AGB yields, and possibly metallicity
effects and star formation efficiencies.  In the Galaxy, interpretation of the 
metal-poor stellar abundances suggest that s-process contributions do not 
occur until [Fe/H]$ \sim -2$ and are not significant until near [Fe/H] $\sim -1$ 
(e.g., Travaglio \etal 2004).   The lower star formation efficiencies in 
dwarf galaxies (Tolstoy \etal 2003; Lanfranchi \& Matteucci 2004; Pagel \& Tautvaisiene 1998) 
mean that metals build up more slowly {\it with time} than in the Galaxy, and thus we might expect 
more contributions from metal-poor stars (e.g., metal-poor AGB stars) at a given 
time or metallicity.   Coupled with the fact that stellar yields are thought to be 
metallicity dependent, then the s- and r-process ratios in stars in dwarf galaxies
may be different from comparable metallicity stars in the Galaxy.  Venn \etal 
(2004) showed that metal-poor stars in dSph galaxies often have lower 
[Y/Eu] and higher [Ba/Y] ratios than similar stars in the Galaxy, consistent 
with metallicity dependent yields by Travaglio \etal (2004).  Thus, the neutron capture element 
ratios offer another tool to distinguish variations in the dSph abundance ratios 
compared with the field star population and possibly the GCs.

In Figure~5, the ratios of [Eu/Fe], [Y/Eu], and [LaBa/Eu] (an average of La and 
Ba, which have similar nucleosynthetic histories and s-process contributions 
in the solar system) are shown.  The Eu abundance primarily samples the pure 
r-process contributions at all metallicities (its r-process in the Sun is 97\%; 
Burris \etal 2000).  Comparing the [Eu/Fe] ratios (Fig.~5, upper panel) to 
the \afe ratios (Fig.~4, upper panel), they follow similar trends as the field stars
as a function of [Fe/H].  This agrees with the idea that $\alpha$-elements and Eu derive 
from the same environment (SNe~II's).  To examine the s-process and r-process contributions separately,
we examine Y (its s-process in the Sun is 74\%; Travaglio \etal 2004), La (its s-process in 
the Sun is 75\%; Burris \etal 2000) and Ba (its s-process in the Sun is 
81\%; Arlandini \etal 1999) relative to Eu.  
Y versus La and Ba are important because 
they sample different peaks in the neutron magic numbers.  Y (Z=39) belongs to 
the first peak that builds through rapid captures around neutron magic number N=50.   
La (Z=57) and Ba (Z=56) belong to the second peak that builds around N=82. 

In the most metal-poor stars ([Fe/H] $< -2$), the observed variations 
in [Eu/Fe] in both the Galactic field stars and GCs has been interpreted as 
variations in the r-process contributions, probably due to inhomogeneous 
mixing due to the location of the forming GC relative to a SNe II event 
(possibly even a SNe II event from the formation of the GC itself) in 
the early chemical evolution of the Galaxy (McWilliam 1997).  The [Y/Eu] 
and [LaBa/Eu] ratios are in excellent agreement with the Galactic field stars, 
suggesting similar r-process contributions of these elements to GCs and field 
stars at low metallicities plus no s-process contributions.  Around [Fe/H]$=-2.0$, 
the [LaBa/Eu] ratio are noticably above the pure $r$-process level ([Ba/Eu]$=-0.7$; 
Burris \etal 2000).  Johnson \& Bolte (2001) argued that this rise may be due 
to problems with the analysis of the strong Ba {\sc ii} 4554 \AA\, line and not 
due to $s$-process contributions.

\begin{figure*}
 \centerline{\epsfig{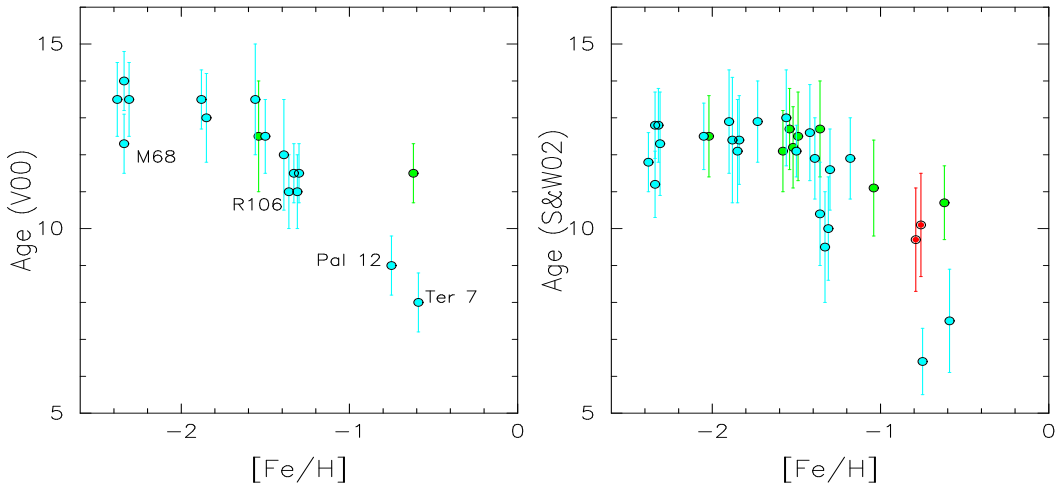}}
 \caption{Globular cluster ages versus [Fe/H] from Vandenberg (2000;
left panel) and Salaris \& Weiss (2002; right panel); same colors used
to distinguish kinematic populations as in Fig.~1.  Leaving out the
young Sagittarius dwarf galaxy globular clusters, both panels show a
trend of younger age with higher metallicity.}
 \label{Fig07}
\end{figure*}

At intermediate metallicities, $-2 <$ [Fe/H] $< -1$, M80 has a slightly high 
value of [Eu/Fe], and Rup~106 has a slightly low ratio,
though Rup~106 is within the range of the halo field stars.  Given that the M80 
data comes from a large sample of stars from a recent study, this result likely 
means M80 was contaminated by higher mass SNe II events.  
M4 has a slightly elevated [LaBa/Eu] ratio.  This is 
likely to be related to a slightly larger AGB s-process contribution than the 
other GCs, however it is not outside the range of the Galactic field stars.   
Ivans \etal (1999) also noted that M4 has slightly higher [Ba/Fe] and [La/Fe] 
ratios compared to other Galactic GCs, and commented ``that the high Ba and 
La properties of M4 stars is surely a primordial, not an evolutionary, effect.''

Among the metal-rich stars and clusters, [Fe/H] $>-1$, the two GCs 
associated with the Sgr remnant, Pal~12 and Ter~7, stand out with slightly 
high values of [Eu/Fe] relative to thick and thin disk stars.  The [Y/Eu] ratio 
in Pal~12 is very low though;  this may suggest a lack of s-process contributions 
from AGB stars, however [LaBa/Eu] is not lacking, and therefore the [Ba/Y] 
ratio in Pal~12 is remarkably high in Figure~6, where we compare the GCs, field 
stars, and dSph field stars.  This is consistent with the [Y/Eu] and [Ba/Y] ratios 
seen in the dSph field stars (Shetrone \etal 2003; Venn \etal 2004), and interpreted in 
terms of more significant contributions from {\it metal-poor} AGB stars in the 
dwarf galaxies.  Metallicity dependent yields from AGB stars (Travaglio \etal 
1999, 2004) suggest that the first-peak r-process elements (including Y, 
Sr, and Zr) are by-passed in favor of second-peak elements (including Ba and La;
and third peak elements as well, but these are not observable in red giant 
spectra) in metal-poor stars, due to an excess of neutrons available for capture 
per seed nucleus.    

Finally, we noted that in the dSph galaxies, a third of the dSph field stars 
with ${\rm [Fe/H]} < -1.8$ show a clear offset in the [Ba/Y] ratios relative 
to the Galactic field stars (Venn \etal 2004).  The metal-poor Galactic GCs do 
not clearly show the same neutron capture element ratios as in the low mass dSph field stars 
(see middle panel of Fig.~6).  It is interesting that comparing the halo and thick disk 
clusters, the halo GCs are offset more toward higher [Ba/Y] similar to dSph field 
stars.

\section{Globular Cluster Ages} 

Given that GCs are thought to be the first objects to form in a galaxy, it 
is very important to determine their true ages.  However, this has been a 
very challenging task (see review by Sarajedini, Chaboyer, \& Demarque 1997).  
While the relative ages between GCs is an easier 
property to measure, different methods have yielded different absolute ages 
for the GCs (e.g., Salaris \& Weiss 2002).

For our purposes, we do not need the absolute ages of the GCs in our sample 
to be the true ages of the GCs.  Given that the WMAP results (Spergel \etal 
2003) find the age of the Universe as $13.7\pm0.2$~Gyr, most of the absolute 
GC ages in the literature are too old.  Here, we only investigate whether the 
GCs show any interesting properties or trends in relative age.   In Figure~7, we show 
age as a function of [Fe/H], where the ages have been determined by Vandenberg (2000) 
and Salaris \& Weiss (2002).  Although Rosenberg \etal (1999) have a sample of relative 
GC ages, we chose to use the more recent results of Salaris \& Weiss (2002) which also 
includes a larger number of GCs.  Vandenberg used stellar evolution models which 
take into account the $\alpha$ element abundance to measure the $V$ magnitude difference 
between the horizontal branch and the main-sequence turn-off ($\Delta 
V^{\rm HB}_{\rm TO}$) for the GCs.  The \afe ratio was assumed to either be 0.3 
or 0.0.  We contrast these ``$\alpha$ enhanced'' isochrone ages with those from 
Salaris \& Weiss, who divide the GCs into four groups according to their 
metallicities then assign ages for a reference cluster within each group using 
$\Delta V^{\rm HB}_{\rm TO}$.  The ages for the remaining GCs in each group are 
then found using isochrones (assuming \afe$=0.4$) to determine the color 
difference between the main sequence turn-off and the red giant branch.  
Of the 45 GCs in our database, 17 and 32 have ages from the Vandenberg 
and Salaris \& Weiss studies, respectively.  We have not searched the literature 
to find absolute ages for the remaining GCs in our sample because of the 
difficulties in normalizing those ages to these two studies.

Comparing the age vs.\ metallicity diagrams in Fig.~7, both studies show  a 
trend of decreasing age with increasing metallicity.  
The most metal-poor thick disk GCs have ages similar to those found in the halo.  
The thin disk clusters show a younger age consistent with the idea that the 
thin disk formed after the halo.  Both methods yield the same (old) age for 
the more metal-poor GCs.  The metal-rich GCs associated with the Sgr remnant, 
Pal~12 and Ter~7, are significantly younger than the Galactic GCs, regardless 
of the differences in \afe adopted between the two studies; i.e., Vandenberg 
(2000) adopted \afe$=0.0$ and Salaris \& Weiss (2002) used \afe$=0.4$.  We note 
that M68, which is amongst the most metal-poor GCs, appears to be slightly 
younger (0.5 to 1 Gyr) than the others.  Whether or not this difference is 
significant will be discussed in \S7.6.  Thus, both methods yield broadly 
similar results for the relative ages of the GCs, however Rup~106 is a good 
example of the difference between the two methods; Salaris \& Weiss find that 
it is younger than the majority of Galactic GCs, whereas Vandenberg finds
that it is not much different in age than the other intermediate metallicity GCs.  

An updated study of relative GC ages by De~Angeli \etal (2005) has also found there there 
is a trend of decreasing age with increasing metallicity.  The metal-poor 
GCs ([Fe/H]$<-1.7$) are shown to be old and coeval with the exceptions of 
M68 and M15.  De~Angeli \etal found that intermediate metallicity GCs are typically 
younger than the metal-poor clusters, but there is a significant spread in the 
ages for these clusters.  While there does appear to be a metallicity-age trend, 
there does not appear to be a $\alpha$-age trend as first noted by Carney (1996).  
If we exclude the extragalactic GCs Rup~106, Pal~12, and Ter~7 from Fig.~4, there 
is very little scatter among the GCs in \afe.  Therefore, we can conclude 
that there is no trend of \afe with age from the existing data.  Given the 
relatively small numbers of metal-rich GCs with abundance ratios, more 
observations of these types of clusters should be done before firm conclusions 
can be drawn about any possible $\alpha$-age relation.

NGC~362 is shown to have a younger age than typical halo GC by Salaris \& 
Weiss (2002).  This reaffirms the previous finding that NGC~362 is younger 
than its second-parameter pair cluster NGC~288 (Catelan \etal 2001).  When the 
horizontal branch of a GC is redder or bluer than expected for its metallicity, 
a ``second-parameter'' is thought to be affecting its morphology.  While it 
is now thought that there are many different parameters that may affect the 
horizontal branch morphology, age is considered the most common parameter 
(e.g., Fusi Pecci \& Bellazzini 1997), as appears to be the case for NGC~362.  
We make note of this because even though it has a younger age, we see no 
peculiarities in the abundance ratios of this GC. 

We want to clarify that although we discuss the difference in ages between 
the clusters, we do not attempt to break the GCs into ``old'' or ``young'' 
halo groups.  Zinn (1993, 1996) looked at the properties of globular clusters 
and separated them into ``young'' and ``old'' halo and thick disk components 
based on the metallicity of the cluster and its horizontal branch morphology.  
Mackey \& Gilmore (2004) have updated Zinn's study to include the most recent 
measurements for Galactic GCs and their own measurements of extragalactic 
GCs, separating them into bulge/disk, ``young'' halo, and ``old'' 
halo components.  In column (9) of Table~3 we list the Mackey \& Gilmore 
classifications for the GCs in our sample.  Mackey \& Gilmore also used the 
Salaris \& Weiss (2002) ages to discuss which GCs are truly young or old.  
Although some of their ``young'' GCs have similar ages to the oldest clusters 
and some ``old'' GCs have relatively young ages, Mackey \& Gilmore classified the 
GCs solely on their [Fe/H] values and horizontal branch morphologies.  We chose 
not to use this classification scheme because there may be other parameters 
beyond metallicity and age affecting a cluster's horizontal branch morphology 
and as a result its classification in Zinn's scheme.

\section{Individual Clusters}

\subsection{The Sagittarius Globular Clusters}

Several GCs have been associated kinematically with the Sgr dwarf galaxy 
merger remnant, including Ter~7, Ter~8, M54, and Arp~2 (Ibata, 
Gilmore, \& Irwin 1994), and Pal~12 (Irwin 1999; Dinescu \etal 2000).   
NGC~2419 was included as a possible member due to its location near 
an overdensity of stars associated with the Sgr tidal tails (Newberg 
\etal 2003).  More recently, NGC~4147 has also been associated with the 
Sgr remnant kinematically (Bellazzini \etal 2003), however this has been 
questioned by Martinez-Delgado \etal (2004) based on an update of the proper 
motion studies by Wang \etal (2000).  
 
M54 and NGC~2419 show no distinctive abundance ratios from the 
Galactic field stars, with the possible exception that NGC~2419 
has a slightly lower [Ca/Fe] than seen in the field.  However, that 
is based on only one star (Shetrone, C\^{o}t\'{e}, \& Sargent 2001).  
Smecker-Hane \& Mcwilliam (2005) found that their three metal-poor 
Sgr field stars have similar \afe abundance 
ratios as stars in the Galactic halo, although Bonifacio \etal (2004) 
suggest that their metal-poor stars may belong to M54.  Unfortunately, 
the M54 analysis is based on only five stars studied by Brown, Wallerstein, 
\& Gonzalez (1999) and at least one of the Smecker-Hane \& McWilliam (2005) 
Sgr stars does have the same metallicity 
and \afe ratio as M54.  Furthermore, Layden \& Sarajedini (1995) suggest 
that M54 may have a significant metallicity spread due to the dispersion 
in the red giant branch.  The other Sgr clusters, Ter~7 and Pal~12, do show 
distinctive kinematic and chemical properties.  Their \afe ratios are lower 
than the typical metal-rich Galactic GC, and more similar to those seen 
in the dSph field stars.   Pal~12 also has low [Y/Eu] and high [Ba/Y] ratios 
like stars in the dSph galaxies.  Both clusters have noticably younger 
ages than other Galactic GC.  If these clusters were not already 
associated with the Sgr remnant, their properties (abundance ratios, kinematics, 
and ages) would set them apart 
from the other Galactic GCs as other studies have found (e.g., 
Tautvai\u{s}ien\.{e} \etal 2004; Cohen 2004; Sbordone \etal 2005).
It has also been shown by these studies that the Sgr GCs follow the abundance 
trends seen in the Sgr field stars.

\subsection{Canis Major Globular Clusters}

Martin \etal (2004) discovery of an overdensity of stars suggested to
be the remnant of a dwarf galaxy in CMa is also associated with four GCs.  
These are NGC~1851, NGC~1904 (M79), NGC~2298, and NGC~2808.   However, we 
note that Pe\~{n}arrubia \etal (2005) used the proper motions of NGC~1851, 
NGC~1904, and NGC~2298 to show that they are not associated with the CMa dwarf.  
Of these GCs, only M79 and NGC~2298 have abundance ratios in the literature.  
Neither of these GCs stands out in a significant way in Fig.~3 
and Fig.~4, although, as seen in Table~1, the data for these two clusters 
is sparce and from older studies.  M79 does have a slightly lower [Si/Fe] ratio, 
but the remaining $\alpha$-element ratios are consistent with field stars of 
similar metallicity.  However, we note that Pe\~{n}arrubia \etal (2005) used 
the proper motions of NGC~1851, NGC~1904, and NGC~2298 to show that they 
are not associated with the CMa dwarf.

Martin \etal also calculated the orbit 
of the CMa galaxy for a prograde or a retrograde orbit.  With the 
knowledge that it is in a prograde orbit, we can also examine the 
clusters near the CMa dwarf galaxy in phase space (see Table~1, Martin 
\etal 2004).  Of the GCs associated with a prograde orbit, only NGC~6205 
(M13), NGC~7078 (M15), NGC~6341 (M92), NGC~4590 (M68), and Rup~106 are 
included in our survey.  Of these GCs, only M68 and Rup~106 stand out.  
Rup~106 is discussed in \S7.3, since it is thought to actually 
have been captured from the Magellanic Clouds.  The properties of M68 
and its possible membership to the CMa dwarf are discussed further in \S7.6.

\subsection{Ruprecht~106}

Rup~106 has long been noted as an interesting cluster given that 
the metallicity from its red giant branch does not match the metallicity 
from the stellar abundances.  For example, Brown, Wallerstein, \& 
Zucker (1997) found [Fe/H]=$-1.45 \pm$0.10 (for our study we use [Fe/H]$=-1.36$, 
which is an average of [FeI/H] and [FeII/H]) 
from spectra of two stars, while Buonanno \etal (1990) found 
${\rm [Fe/H]}=-1.9\pm0.2$ from the red giant branch.  Buonanno \etal 
also found that Rup~106 has a younger age than the typical halo GC 
(see Fig.~7).  Lin \& Richer (1992) speculated that Rup~106 was acquired 
by our Galaxy from the Magellanic Clouds based on its position and 
assorted other odd properties for an outer halo cluster (young age, 
high number of blue stragglers, and horizontal branch morphology)
Rup~106 has a radial velocity of $-44 \pm 3$ \kms (Harris 1996), but no other 
kinematics exist.  As discussed in \S3, we examine potential
orbits for Rup~106 from reasonable limits on its (undetected)
proper motion to find that it has either a typical halo orbit or
a plunging orbit.
 
The $\alpha$-element ratios for Rup~106 are significantly lower than the Galactic
field stars of similar metallicity, which is similar to the stars 
in the dSph galaxies (see Figs.~3 and 4).   We also notice that [Eu/Fe] is lower than 
in most of the GCs and Galactic field stars (see Fig.~5), also suggesting that 
Rup~106 has had less SNe~II r-process contributions. 

If Rup~106 had its origin in the Magellanic Clouds, then we can
compare its abundance ratios with other LMC GCs.  Hill \etal (2000) and 
Hill (2004) found that the LMC GCs have [Ca/Fe] and [Ti/Fe] ratios below 
what is seen in the Galactic halo stars at all metallicities.  Rup~106 falls near these 
LMC GCs abundances, but typically with abundance ratios about 0.1~dex lower.  
Johnson \etal (2004) also determined the abundance ratios for three old 
LMC GCs.  They find the [Ca/Fe] ratio for Hodge~11 to be similar to the 
Galactic field stars, which is much higher than the other two LMC GCs 
in their sample and those in the Hill study.  The [Ti/Fe] ratios are 
similar in both studies.  While the abundance 
ratios in Rup~106 do not match precisely with the analysed GCs, the tendancy 
towards lower \afe is more consistent with the LMC clusters and dSph field stars.

\subsection{Bulge Globular Clusters} 

As noted in \S3, we classified anything within $R_{GC}\le2.7$~kpc to be within 
the bulge, unless the kinematics associate the GC with a different Galactic component.  
For example, Dinescu \etal (2003) noted that NGC~6528 and NGC~6553 are confined to the 
bulge region, and suggest that NGC~6528 is actually a ``genuine Milky Way 
bar cluster.''  Although NGC~6553 is within our bulge limit, its kinematics show 
it has a circular orbit and therefore it is classified as a thin disk cluster.   

From our model, we classify nine clusters as belonging to the bulge.  However, only 
one of these clusters (NGC~6528) has known phase space velocities.  The bulge 
stars have been shown to be mostly metal-rich ([Fe/H]$\ge-1.6$; McWilliam \& Rich 
2004).  Using this limit, six of the nine GCs fall in this range with Ter~4 being 
right at this limit.  The three clusters NGC~6287, NGC~6293, and NGC~6541 are more 
metal-poor than this ([Fe/H]$<-1.8$) and may not be truly associated with the bulge.

The bulge stars are known to show elevated \afe ratios compared to the metal-rich 
thin disk stars (McWilliam \& Rich 2004; Fulbright, Rich, \& McWilliam 2004).  
In general, the bulge GCs show a similar trend, but not as clear as seen in the 
bulge field stars.  Of the $\alpha$ elements, [Ca/Fe] shows the clearest difference 
between the bulge GCs and the thin disk stars.  This is somewhat contrary 
to the bulge field stars.  They have been found to be more enhanced compared to 
metal-rich thin disk stars in [Mg/Fe] and [Si/Fe] than [Ca/Fe] and [Ti/Fe] 
(McWilliam \& Rich 2004).  McWilliam \& Rich suggested that this implied 
SNe II played a greater role in the bulge than in the thin disk.

\begin{figure*}
 \centerline{\epsfig{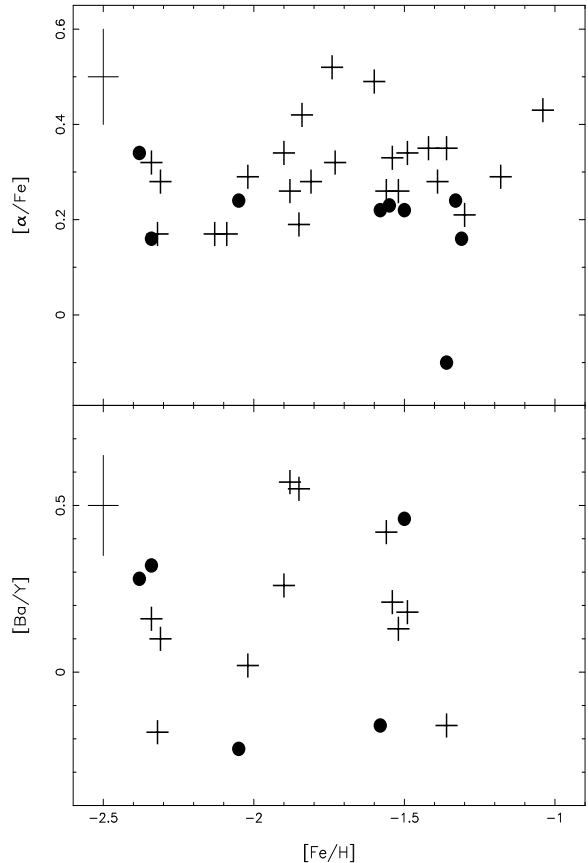}}
 \caption{A comparison of ``young'' (filled circles) and ``old'' (plusses)
halo GCs as defined by Mackey \& Gilmore (2004) using \afe and [Ba/Y] as
a function of [Fe/H].  Mackey \& Gilmore argue that all the ``young''
halo GCs have been captured from dwarf galaxies.  Although the ``young''
GCs appear to have marginally lower \afe ratios, the difference between
them and the ``old'' halo GCs is not significant.  The lone ``young'' GC with low
\afe is Rup~106, which is discussed in \S7.3.  The error bars represent
a mean uncertainty of $\pm0.10$ for [$\alpha$/Fe], $\pm0.15$ for [Ba/Y],
and $\pm0.05$ for [Fe/H]}
 \label{Fig08}
\end{figure*}

\subsection{Thick Disk Clusters}

As noted in \S3, it is interesting to see the majority of thick disk GCs 
clustered toward the metal-poor range of the thick disk field stars.  
Of the ten thick disk clusters, seven have $\Pi,\Theta,W$ velocities, 
which makes their classifications more certain since it is based on 
phase space sampling (not just physical location).  As discussed in \S6, the 
metal-poor thick disk GCs have similar ages to halo clusters, while the 
two metal-rich ([Fe/H]$>-1.0$) GCs show a younger age (see Fig.~7).  The implication 
is that when discussing the formation of the thick disk, it must be taken into 
account that the metal-poor thick disk GCs formed early on and with similar 
abundance ratios as the metal-poor thick disk field stars.  
Thus, the metal-poor thick disk represents the transitional phase between the 
halo and thick disk formation as the Galaxy collapsed according to the 
Eggen, Lynden-Bell \& Sandage (1962) model of galaxy formation.  Examining the 
M31 GCs, Morrison \etal (2004) found that 40\% have disk kinematics 
and are metal-poor (with no age estimates), which is similar to our findings.  
Contrary to this, a recent study by Mould (2005) found that the thick disks in four edge-on 
galaxies are old and relatively {\it metal-rich}.  This suggests that 
thick disks form relatively early on and from gas-rich mergers (Dalcanton 
\& Bernstein 2002; Brook \etal 2005).  Further studies of the Galactic thick 
disk field stars and GCs need to be done in order to resolve the formation of 
the thick disk.

\subsection{Miscellaneous Clusters: M68 and $\omega$~Centauri}

{\it NGC~4590 (M68):}
M68 stands out from the majority of the Galactic GCs,
primarily due to its high Galactocentric rotational velocity ($\Theta = 
304$ \kms).  As noted by Lee, Carney, \& Habgood (2004), M68 also has 
some interesting abundance ratios.  As seen in Fig.~3, the [Ti/Fe] ratio 
is lower than other GCs and field stars of similar metallicites.  On 
the other hand, the [Si/Fe] ratio in M68 is unusually high.  
M68 is also interesting because its horizontal branch is 
notably redder than expected for a GC of its metallicity (Harris 1996).  
This is an example of the second-parameter effect as discussed in 
\S6.  As shown in Fig.~7, M68 is thought to have a slightly younger age 
($12.3\pm0.8$ Gyr, Vandenberg 2000; see also De~Angeli \etal 2005) compared 
to the typical halo GC ($13.2\pm0.9$ Gyr).  This age difference is not large 
enough to seriously affect the horizontal branch morpology though 
according to theoretical models (e.g., Catelan \& de Freitas Pacheco 1993; 
Lee, Lee, \& Gibson 2002), and argues that age alone cannot be the sole 
parameter affecting the horizontal branch in M68.  

Given the slightly younger age of M68, its high rotational velocity, 
and its unique [Si/Ti] ratio, we argue that it is a candidate extragalactic 
GC.  Martin \etal (2004) included M68 as a possible CMa dwarf cluster.  
However, the membership of certain GCs with the CMa dwarf has been called 
into question by Pe\~{n}arrubia \etal (2005).  Using Fig.~11 of Pe\~{n}arrubia 
\etal and the kinematics for M68 (Dinescu 2004, private communications), 
it seems unlikely that M68 belongs to the CMa dwarf. 

{\it $\omega$ Centauri:}
We briefly comment on $\omega$~Centauri ($\omega$~Cen).
This cluster has long been known to have a spread in its metallicity 
(Dickens \& Woolley 1967).  Because of this spread, $\omega$~Cen has 
been suggested to have once been the nucleus of a now accreted dwarf 
galaxy (e.g., Rey \etal 2004).  Using the kinematic data from Dinescu 
\etal (1999b, and private communications for updates), $\omega$~Cen is 
a typical halo GC ($\Pi,\Theta,W=-26.9,
-71.0,-10.7$).  Pancino \etal (2002) and Origlia \etal 
(2003) obtained spectra for the different populations in $\omega$~Cen 
including the recently discovered high metallicity population (Pancino
\etal 2000).  Both studies have found abundance ratios for [O/Fe], 
[Si/Fe], [Mg/Fe], and [Ca/Fe] from ${\rm [Fe/H]}\sim-1.6$ to $\sim-1.0$ 
that are consistent with the Galactic field stars, i.e., halo field star 
ratios, with a bend towards lower abundance ratios at higher metallicities 
(${\rm [Fe/H]}>-1.0$).  In addition, Pancino (2004) presented the preliminary 
[Ca/Fe] results of about 700 red 
giants in $\omega$~Cen which illustrated the ``turn" toward lower 
[Ca/Fe] for the metal-rich stars.  This shows that the star formation 
in $\omega$~Cen took place over an extended period of time and that 
it was enriched in a similar manner as seen in galaxies.

\begin{figure*}
 \centerline{\epsfig{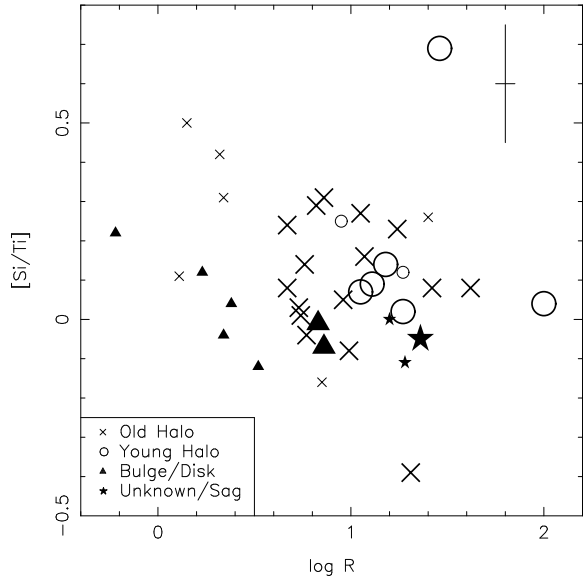}}
 \caption{[Si/Ti] plotted as a function of distance for GCs as classified
by Mackey \& Gilmore (2004; see column [9] of Table~3).  The larger symbols are for
those GCs plotted at their apogalactic distance (Dinescu 2004, private communications),
while the others are at their Galactocentric radius.  We find that there
is no clear trend of decreasing [Si/Ti] with increasing $\log\,R$ for
the ``old'' halo population.  The error bars represent a mean uncertainty of
$\pm0.15$ for [Si/Ti] and $\pm0.05$ for [Fe/H], although individual clusters
may have smaller errors.}
 \label{Fig09}
\end{figure*}

\section{Discussion}

\subsection{Globular Clusters and Merger Events}

Out of the 45 GCs with high-resolution abundances, four GCs are excellent 
candidates for having an extragalactic origin (Rup~106, Ter~7, Pal~12, and M68).  This 
is approximately one-eleventh of our sample.  Two of the four candidate 
extragalactic GCs belong to the Sgr remnant, Ter~7 and Pal~12. 
Rup~106 has been suggested as being captured from the Magellanic Clouds or 
as part of the CMa remnant (prograde orbit).  
We have included M68 as a possible extragalactic cluster due to its unusual 
kinematics, its younger age, and its unique [Si/Ti] ratio.
The detection of extragalactic metal-rich GCs seems relatively easy (e.g., Pal~12 
and Ter~7) because the \afe ratios are clearly distinctive.  The influence of 
Type Ia SN occuring (presumably) at a lower metallicity in the chemical evolution 
of a dwarf galaxy causes the ``turn" towards lower \afe at lower 
metallicities than in the Galactic field stars leading to clear differences between
their stellar populations.  

Mackey \& Gilmore (2004) analyzed the physical properties of GCs in the Milky Way and its 
dwarf companions to allow a comparison between their GC systems.  They divided up 
the Galactic GCs according to their metallicities and horizontal branch morphologies 
into bulge/disk, ``young'' halo, and ``old'' halo components, which we list in column (9) of 
Table~3.  Overall, our classifications match up well with those from Mackey \& Gilmore 
with the exception of a few cases where the kinematics tend to place a cluster in 
the thick disk while they classify it as belonging to the halo.  
Mackey \& Gilmore concluded that all of the ``young'' halo clusters and a small fraction 
(15-17\%) of ``old'' halo clusters have been captured from merged dwarf galaxies, which 
does not include the Sgr GCs.  This is much higher than what we would estimate from 
the abundance ratios, even including the Sgr GCs.  In Figure~8, we compare the \afe and 
[Ba/Y] ratios for the ``old'' and ``young'' halo GCs as classified by Mackey \& Gilmore.  
Although the \afe ratios in ``young'' halo GCs are marginally lower than the ``old'' halo 
GCs, the offset is not significant, with the exception of Rup~106 which 
is thought to be a captured GC.  

As noted above, the lack of noticable variations between the Galactic GCs and field 
stars in our sample argues against a large number of accretions by dwarf galaxies 
with GCs.  Because of the noticable offsets between the dSph and Galactic field stars 
at metallicities of [Fe/H]$=-1.6$ and higher, for example in the \afe ratios (see Fig.~4), 
it should be relatively easy to pick out metal-rich extragalactic GCs.  However, what 
about metal-poor extragalactic GCs?  Can we rule out early accretions by dwarf galaxies 
before major chemical evolution has occured?  Are there any abundance indicators, such as [Ba/Y] 
(Venn \etal 2004), that show differences for the metal-poor stars in dwarf 
galaxies and the Galactic field stars?  The more metal-poor Sgr GC M54 does not 
stand out in any way from the Galactic field stars, although it may not be the 
ideal example given that it may be the nucleus of the Sgr dwarf galaxy.  None of the 
candidate clusters associated with the CMa dwarf galaxy stand out in a significant 
way from the rest of the Galactic field according to their abundance ratios either.  
None of the metal-poor GCs in our compilation stand out in [Ba/Y] in the same way that 
the dSph stars do.  However, a comparison between the thick disk and halo GCs at 
intermediate metallicities ([Fe/H]$\sim-1.6$) shows that the halo clusters are more 
like the dSphs stars than the thick disk clusters in that the halo GCs are shifted 
toward higher [Ba/Y] values than the thick disk ones.

One way to resolve this problem is to examine the abundances in known extragalactic 
clusters such as those in the LMC, Fornax, and Sgr.  Many of these clusters have 
been shown to have ages similar to the Galactic GCs (e.g., Buonanno \etal 1998, 1999).  
Therefore, it will be key to determine what similarities and, more importantly, what 
differences there are between metal-poor clusters and Galactic field stars of similar 
metallicities.  For example, as noted in \S7.3, Hill (2004) found that the LMC GCs tend 
to have [Ca/Fe] and [Ti/Fe] ratios that are below those of the Galactic halo stars even 
for the metal-poor ones.  This is similar to what is seen in Rup~106.  For more metal-poor 
GCs, the picture is less clear.  Before firm conclusions can be drawn, more work will 
need to be done on the metal-poor dwarf galaxy stars and their GCs to help determine 
what kind of impact early mergers may have had in the formation of our Galaxy.

\begin{figure*}
 \centerline{\epsfig{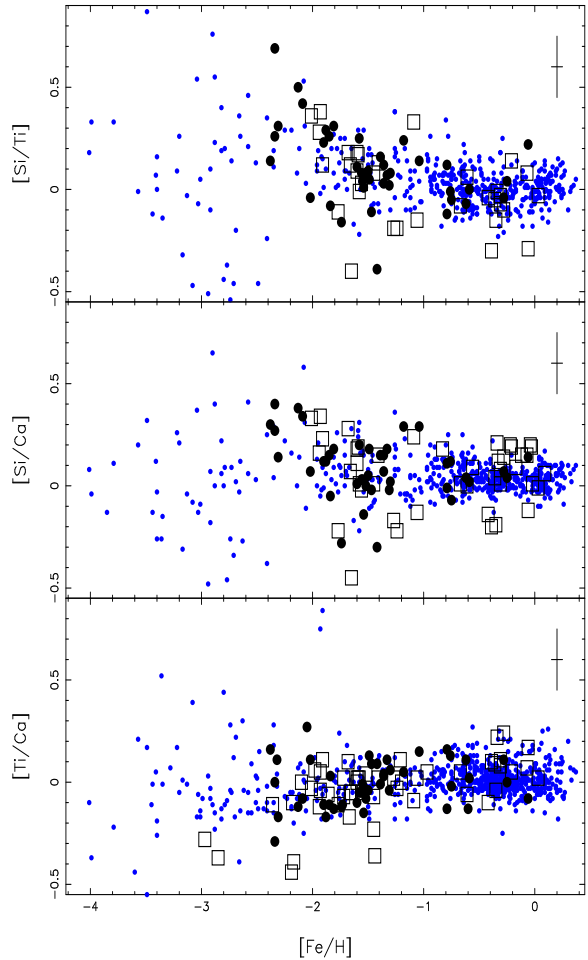}}
 \caption{The [Si/Ti], [Si/Ca], and [Ti/Ca] for the Galactic field stars
(blue filled circles), GC (black filled circles), and dSph stars (open squares) as a function
of [Fe/H].  The scatter in [Si/Ti] and [Si/Ca] for stars with [Fe/H]$<-2.2$,
and the relative lack thereof in [Ti/Ca], shows that these stars formed from ejecta
of stars with a variety of stellar masses.  The following trend of
decreasing [Si/Ti] and [Si/Ca] and then leveling off with increasing [Fe/H]
argues for the idea that more massive stars ($M>30M_{\odot}$) played a
larger role in the early formation of the Galaxy until the interstellar
medium became integrated and less massive stars contributed.  See further discussion
in \S8.2.  The error bars represent a mean uncertainty of $\pm0.15$ for relative
$\alpha$-element ratios and $\pm0.05$ for [Fe/H]}
 \label{Fig10}
\end{figure*}

\subsection{Early Chemical Evolution}

Lee \& Carney (2002) discussed a trend of decreasing [Si/Ti] ratios with 
increasing Galactocentric radius ($R_{GC}$) for the ``old'' halo GCs as classified 
by Zinn (1993).  They argued that this gradient could be explained by differences 
in SNe II contributions by stellar progenitors with different masses according to 
models by Woosley \& Weaver (1995).  The models suggest that more massive stars ($M>30M_{\odot}$)
yield higher amounts of silicon relative to calcium and titanium.  Therefore, 
Lee \& Carney reasoned that the central regions of the Galaxy, with a deeper 
gravitational potential, retained a higher amount of the more massive SNe II 
ejecta than the outer regions of the Galaxy.  Given our larger sample of GCs, we 
plot [Si/Ti] against $\log\,R$ in Figure~9, where the larger symbols are for GCs plotted 
at their apogalactic distance (Dinescu 2004, private communications) and the smaller 
ones are those at their Galactocentric distance.  To be consistent with the 
Lee \& Carney analysis, we have plotted the clusters according to their classification 
in Mackey \& Gilmore (2004).  Mean errors are adopted for [Si/Ti] and $\log\,R$ 
(the true erros for some of the individual GCs may be smaller than these).  
Excluding M79 whose ratios come from an older study (Gratton \& Ortolani 1989), 
there is no apparent trend of decreasing [Si/Ti] with increasing $R$.  

In investigating [Si/Ti], we noticed that all the GCs with [Si/Ti]$>0.4$ 
have [Fe/H]$<-1.90$.  To examine this further, [Si/Ti], [Si/Ca], and 
[Ti/Ca] are plotted as a function of [Fe/H] in Figure~10.  For the stars with [Fe/H]$<-2.4$, 
[Si/Ti] and [Si/Ca] are scattered over a wide range of values, but [Ti/Ca] remains relatively 
flat across all metallicities.  This suggests spatial inhomogeneities where the oldest stars 
were likely polluted by individual SNe II events from progenitors of varying progenitor stellar 
masses (a majority with $M>30M_{\odot}$; Woosley \& Weaver 1995).  Those forming from the 
ejecta of more massive SNe II events are predicted to have higher amounts of silicon than 
calcium or titanium, while the calcium and titanium yields are similar over all stellar 
masses.  Shetrone (2004) found a similar trend in the dSph stars, where the [O/Fe] and [Mg/Fe] 
ratios tended to be higher than [Ca/Fe] and [Ti/Fe] for the more metal-poor dSph stars 
([Fe/H]$<-1.7$).  

Between [Fe/H]$\sim-2.4$ to $\sim-1.6$, there appears to be a trend of decreasing [Si/Ti] and 
[Si/Ca].  This is consistent with a decrease in the ejecta of the most massive stars.  
All of the different populations, Galactic field stars, GCs, and dSph stars, show  
this same effect.  Above [Fe/H]$\sim-1.6$, the relations of [Si/Ti] and [Si/Ca] are 
relatively flat for Galactic field stars and GCs, which suggests that the Galactic 
interstellar medium has become fully mixed in ejecta from all SNe II masses.  The two 
GCs with clearly lower [Si/Ti] and [Si/Ca] ratios come from an older study (NGC~4833 
and M79; Gratton \& Ortolani 1989) and should be reinvestigated.

It is interesting to see that as many as one third of the stars in the dSph galaxies 
fall below the majority of the Galactic field stars in both [Si/Ti] and [Si/Ca].  Given 
that these same stars show no differences to the Galactic field stars in [Ti/Ca], 
something must be reducing the silicon abundances.  That the dSph stars in the range 
$-2.2<{\rm [Fe/H]}<-1.6$ are similar to the Galactic field star and the GC abundance ratios
argues against the possibility of a truncated initial mass function in the dSph 
galaxies (discussed by Tolstoy \etal 2003) because the initially higher silicon yields 
presumably came from more massive stars as discussed above.  However, a 
preferential outflow of ejecta from the most massive stars could lead to a reduced amount 
of silicon in the dSph interstellar medium.

\section{Summary}

We have produced a standardized dataset of select abundance ratios for the
Galactic GCs.  We also divided up the GCs into their respective
Galactic components to better compare them to the Galactic field stars.
In general, the Galactic GCs follow the trends seen in the field stars
for both the $\alpha$ and neutron capture elements.  Therefore,
the similarity in the abundance ratios over a wide range of metallicities seems
to indicate that the chemical evolution of the Galaxy has been similar throughout.

We also find that GCs assigned to the thick disk tend to be found at the metal-poor
end of the thick disk field star distribution.  Given the similar ages of these
GCs with the metal-poor halo GCs, this seems to indicate that the metal-poor
population of the thick disk formed in a manner similar to the halo, although some 
differences were seen in [Ba/Y].  On the other hand, the
bulge is a distinct component; McWilliam \& Rich (2004) and Fulbright, Rich, \& 
McWilliam (2004) found that the bulge
stars tend to have enhanced $\alpha$ elements compared to field stars and
we confirm this tendancy for the bulge GCs as well.

From our sample, we have found four extragalactic GC candidates:  Ter~7, Pal~12, 
Rup~106, and M68.  Ter~7 and Pal~12 are both known to be associated with the Sgr 
dwarf galaxy, while Rup~106 is thought to be associated with either the Magellanic 
Clouds or the CMa dwarf galaxy.  M68 is interesting because of its high Galactocentric rotational 
velocity, its slightly younger age, and its low [Ti/Fe] and high [Si/Fe] ratios.  

There is an interesting trend of decreasing [Si/Ti] with increasing [Fe/H] 
for Galactic field stars and GCs in the approximate range of $-2.2<{\rm [Fe/H]}<-1.6$.  
This implies that the most massive stars ($M\ge30M_{\odot}$; Woosley \& Weaver 1995) 
played a larger role during the initial chemical evolution of the Galaxy that lead to 
higher yields of silicon.  The dSph stars also follow this trend, yet the [Si/Ti] 
ratios in about one-third of the sample appear lower than in the Galactic field stars.  
This implies that silicon has been lost in dSph galaxies, presumably by a preferential 
outflow of the ejecta from massive SNe II events.

\acknowledgements
Thank you to the referee B. Carney who gave very valuable comments which aided in clarifying 
our discussions.  B.J.P. and K.A.V. would like to thank the National Science Foundation (NSF) for 
support through a CAREER award, AST 99-84073.  Thanks to D. Dinescu for sharing her 
updated kinematic information for the GCs and for discussions concerning those 
kinematics.  Thank you to J. Cohen, J. Melendez, J.-W. Lee, M. Shetrone, B. Castilho, 
G. Wallerstein, \& G. Tautvai\u{s}ien\.{e} for sharing their solar abundances and 
$\log\,gf$ values.


\clearpage

\LongTables
\begin{deluxetable}{lrl}
\tablewidth{0pc}
\tablecaption{Globular Cluster List and References \label{tbl-1}}
\tablehead{
\colhead{Cluster} & \colhead{No.} & \colhead{Reference}
          }
\startdata
104 (47 Tuc) &  3 & Gratton, Quarta, Ortolani (1986) \\
             &  4 & Brown \& Wallerstein (1992) \\
             &  1 & Norris \& Da Costa (1995)\tablenotetext{a}{Cluster with either no solar abundances and
$\log\,gf$ values given or have values that are similar to the adopted ones.} \\
             & 12 & Carretta \etal (2004)\tablenotemark{a} \\ 
             & 12 & James \etal (2004b)\tablenotemark{a} \\
288          &  2 & Gratton (1987) \\
             & 13 & Shetrone \& Keane (2000)\tablenotemark{a} \\
362          &  1 & Gratton (1987) \\
             & 12 & Shetrone \& Keane (2000)\tablenotemark{a} \\
1904 (M79)   &  2 & Gratton \& Ortolani (1989)\tablenotemark{a} \\
2298         &  3 & McWilliam, Geisler, \& Rich (1992) \\
2419         &  1 & Shetrone, C\^{o}t\'{e}, \& Sargent  (2001)\tablenotemark{a} \\
3201         &  3 & Gratton \& Ortolani (1989)\tablenotemark{a} \\
             & 18 & Gonzalez \& Wallerstein (1998) \\
4590 (M68)   &  2 & Gratton \& Ortolani (1989)\tablenotemark{a} \\
             &  1 & Shetrone \etal (2003)\tablenotemark{a} \\
             &  7 & Lee, Carney, \& Habgood (2004)\tablenotemark{a} \\
4833         &  2 & Gratton \& Ortolani (1989)\tablenotemark{a} \\
5272 (M3)    &  7 & Kraft \etal (1993)\tablenotemark{a} \\
             &  3 & Kraft \etal (1995)\tablenotemark{a} \\
             &  3 & Shetrone, C\^{o}t\'{e}, \& Sargent (2001)\tablenotemark{a} \\
             & 28 & Sneden \etal (2004b)\tablenotemark{a} \\
             & 13 & Cohen \& Melendez (2004a) \\
5466         &  1 & McCarthy \& Nemec (1997)\tablenotemark{a} \\ 
5897         &  2 & Gratton (1987) \\
5904 (M5)    &  3 & Gratton, Quarta, \& Ortolani (1986) \\
             & 13 & Sneden \etal (1992)\tablenotemark{a} \\
             & 28 & Ivans \etal (2001)\tablenotemark{a} \\ 
             & 23 & Ram\'{i}rez \& Cohen (2003) \\
6093 (M80)   & 10 & Cavallo, Suntzeff, \& Pilachowski (2004) \\
6121 (M4)    &  3 & Gratton, Quarta, \& Ortolani (1986) \\
             &  3 & Brown \& Wallerstein (1992) \\
             & 24 & Ivans \etal (1999)\tablenotemark{a} \\
6205 (M13)   & 35 & Kraft \etal (1997)\tablenotemark{a}; Sneden \etal (2004b)\tablenotemark{a} \\
             & 13 & Cohen \& Melendez (2004a) \\
6218 (M12)   &  1 & Mishenina, Panchuk, \& Samus (2003)\tablenotemark{a} \\
6254 (M10)   &  2 & Gratton \& Ortolani (1989)\tablenotemark{a} \\
             & 14 & Kraft \etal (1995)\tablenotemark{a} \\
             &  2 & Mishenina, Panchuk, \& Samus (2003)\tablenotemark{a} \\
6287         &  3 & Lee \& Carney (2002) \\
6293         &  2 & Lee \& Carney (2002) \\
6341 (M92)   &  6 & Shetrone (1996)\tablenotemark{a} \\
             & 30 & Sneden, Pilachowski, \& Kraft (2000a)\tablenotemark{a} \\
             &  4 & Shetrone, C\^{o}t\'{e}, \& Sargent(2001)\tablenotemark{a} \\ 
6342         &  4 & Origlia, Valenti, \& Rich (2005)\tablenotemark{a} \\
6352         &  3 & Gratton (1987) \\
6362         &  2 & Gratton (1987) \\
6397         &  3 & Gratton \& Ortolani (1989)\tablenotemark{a} \\
             &  2 & Norris \& Da Costa (1995)\tablenotemark{a} \\
             & 16 & Castilho \etal (2000)\tablenotemark{a} \\
             &  8 & James \etal (2004b)\tablenotemark{a} \\
6528         &  4 & Carretta \etal (2001) \\
             &  3 & Zocalli \etal (2004) \\
             &  4 & Origlia, Valenti, \& Rich (2005)\tablenotemark{a} \\
6541         &  2 & Lee \& Carney (2002) \\
6553         &  2 & Barbuy \etal (1999) \\
             &  5 & Cohen \etal (1999) \\
6656 (M22)   &  3 & Gratton \& Ortolani (1989)\tablenotemark{a} \\
             &  7 & Brown \& Wallerstein (1992) \\
6715 (M54)   &  5 & Brown, Wallerstein, \& Gonzalez (1999) \\
6752         &  3 & Gratton, Quarta, \& Ortolani (1987) \\
             &  6 & Norris \& Da Costa (1995)\tablenotemark{a} \\
             & 18 & Gratton \etal (2001)\tablenotemark{a} \\
             & 20 & Yong \etal (2003) \\
             & 11 & Cavallo, Suntzeff, \& Pilachowski (2004) \\
             & 18 & James \etal (2004a)\tablenotemark{a} \\
6809 (M55)   &  2 & Shetrone \etal (2003)\tablenotemark{a} \\
6838 (M71)   &  3 & Gratton, Quarta, \& Ortolani (1986) \\
             &  8 & Sneden \etal (1994)\tablenotemark{a} \\
             &  8 & Ram\'{i}rez \& Cohen (2002) \\
             &  2 & Mishenina, Panchuk, \& Samus (2003)\tablenotemark{a} \\ 
             &  2 & Lee, Carney, \& Balachandran (2004)\tablenotemark{a} \\
7006         &  6 & Kraft \etal (1998)\tablenotemark{a} \\
7078 (M15)   & 18 & Sneden \etal (1997)\tablenotemark{a} \\ 
             &  3 & Sneden \etal (2000)\tablenotemark{a} \\
             & 31 & Sneden, Pilachowski, \& Kraft (2000)\tablenotemark{a} \\
7099 (M30)   &  1 & Shetrone \etal (2003)\tablenotemark{a} \\
7492         &  4 & Cohen \& Melendez (2004b) \\
-- (Lil 1)   &  2 & Origlia, Rich, \& Castro (2002)\tablenotemark{a} \\
-- (Pal 5)   &  4 & Smith, Sneden, \& Kraft (2002)\tablenotemark{a} \\
-- (Pal 6)   &  3 & Lee, Carney, \& Balachandran (2004) \\
-- (Pal 12)  &  2 & Brown, Wallerstein, \& Zucker (1997) \\
             &  4 & Cohen (2004) \\
-- (Rup 106) &  2 & Brown, Wallerstein, \& Zucker (1997) \\
-- (Ter 4)   &  4 & Origlia \& Rich (2004)\tablenotemark{a} \\
-- (Ter 5)   &  6 & Origlia \& Rich (2004)\tablenotemark{a} \\
-- (Ter 7)   &  3 & Tautvai\u{s}ien\.{e} \etal (2004)\tablenotemark{a} \\
             &  5 & Sbordone \etal (2005) \\
\enddata
\end{deluxetable}

\clearpage

\begin{figure*}
 \centerline{\epsfig{figure=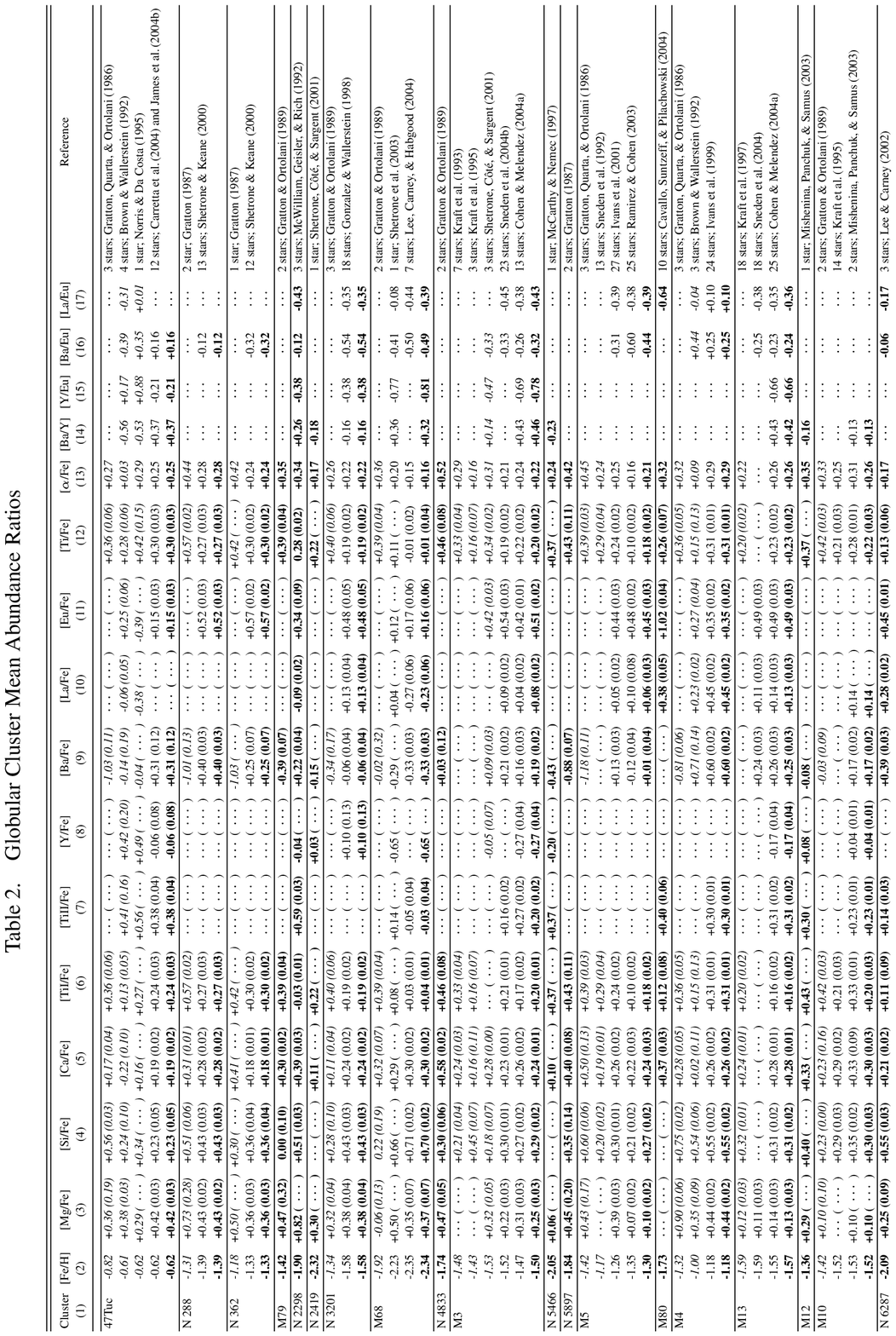}}
 \label{table2p1}
\end{figure*}

\begin{figure*}
 \centerline{\epsfig{figure=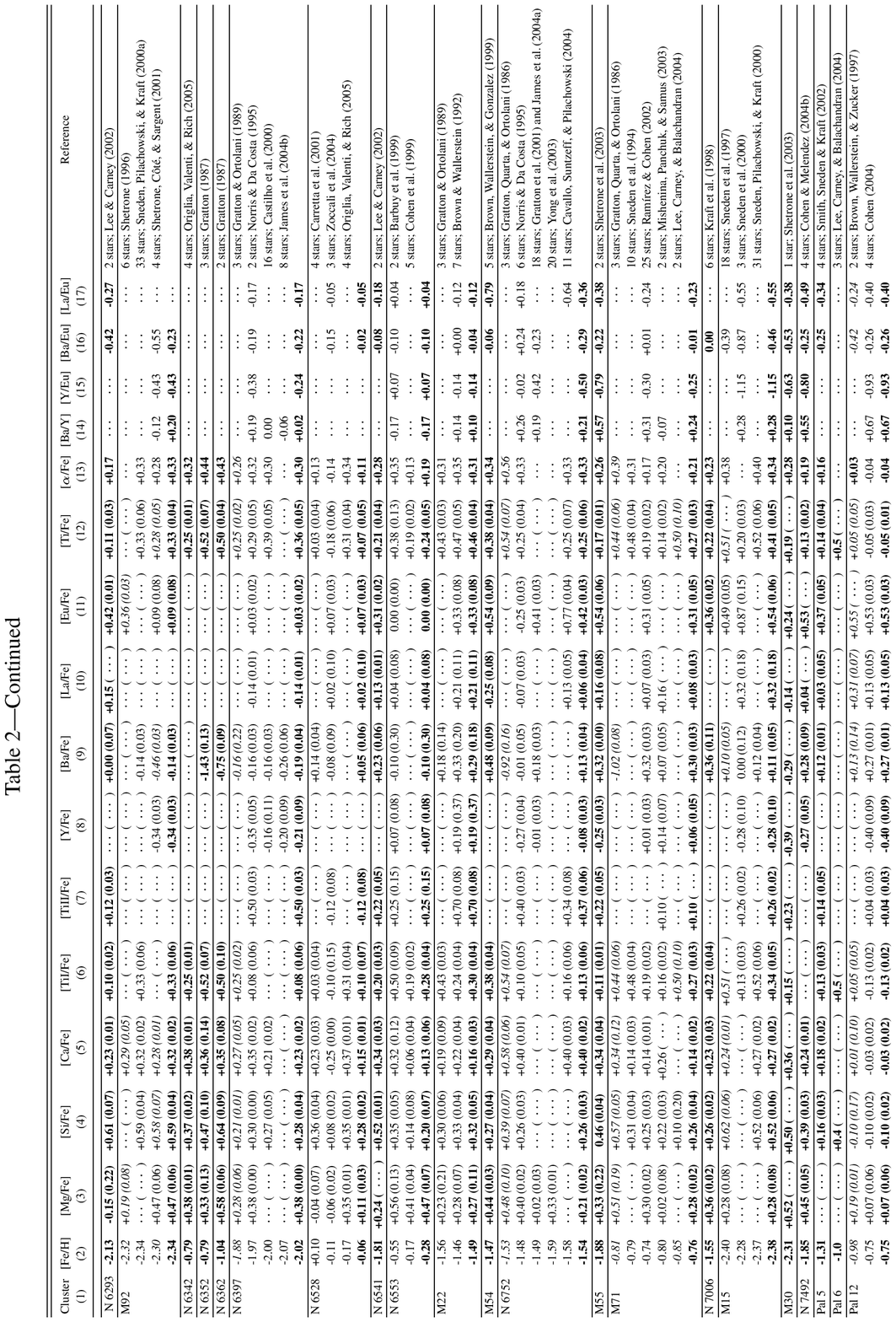}}
 \label{table2p2}
\end{figure*}

\begin{figure*}
 \centerline{\epsfig{figure=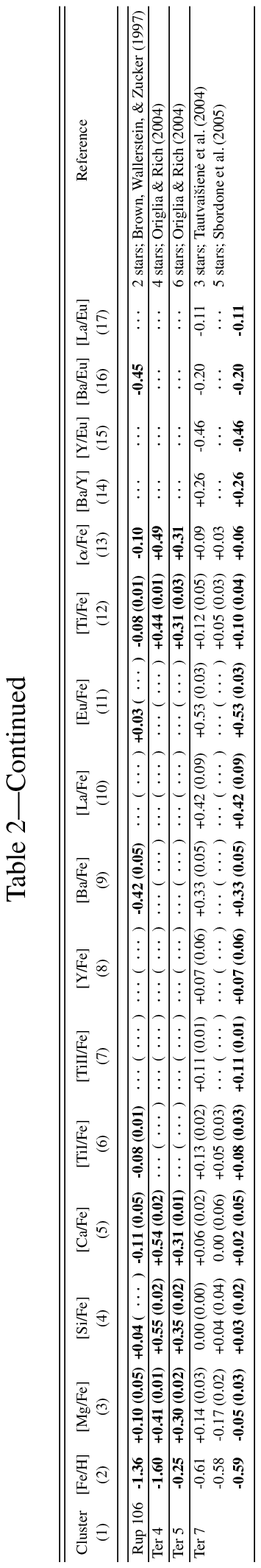}}
 \label{table2p3}
\end{figure*}

\clearpage

\tabletypesize{\footnotesize} 
\begin{deluxetable}{lcccccccccccc}
\tablewidth{0pc}
\tablecaption{Globular Cluster Proper Motions \label{tbl-3}}
\tablenum{3}
\tablehead{
\colhead{Cluster} & \colhead{X} & \colhead{Y} & \colhead{Z} & \colhead{$\Pi$} & 
\colhead {$\sigma_\Pi$} & \colhead{$\Theta$} & \colhead{$\sigma_\Theta$} & 
\colhead{$W$} & \colhead{$\sigma_W$} &
\colhead{Class1} & \colhead{Class2} \\
\colhead{(1)} & \colhead{(2)} & \colhead{(3)} & \colhead{(4)} &
\colhead{(5)} & \colhead{(6)} & \colhead{(7)} & \colhead{(8)} & 
\colhead{(9)} & \colhead{(10)} & \colhead{(11)} & \colhead{(12)} 
          }
\startdata
47 Tucanae   &     6.63 &   -2.58 &   -3.18 &     19.4 &    10.2 &   177.1 &     7.8 &    40.3 &     3.6 &  TK & BD \\
NGC 288      &     8.58 &    0.04 &   -8.80 &     18.7 &     9.5 &   -50.5 &    20.1 &    52.8 &     0.4 &   H & OH \\
NGC 362      &     5.43 &   -5.01 &   -6.14 &     81.4 &    28.4 &   -20.2 &    28.5 &   -79.0 &    22.1 &   H & YH \\
M79          &    16.14 &   -8.25 &   -6.32 &     93.8 &    31.2 &    81.2 &    30.4 &    12.4 &    35.8 &   H & OH \\
NGC 2298     &    12.74 &   -9.37 &   -2.95 &    -61.7 &    40.6 &   -49.5 &    34.8 &   116.5 &    51.0 &   H & OH \\
NGC 2419     &    84.66 &   -0.49 &   35.90 &  \nodata & \nodata & \nodata & \nodata & \nodata & \nodata &   H & OH \\
NGC 3201     &     7.88 &   -4.90 &    0.75 &  \nodata & \nodata & \nodata & \nodata & \nodata & \nodata &  TK & YH \\
M68          &     4.42 &   -7.17 &    6.00 &   -115.3 &    27.7 &   303.5 &    34.8 &    13.3 &    25.0 &   H & YH \\
NGC 4833     &     4.94 &   -5.36 &   -0.91 &  \nodata & \nodata & \nodata & \nodata & \nodata & \nodata &  TK & OH \\
M3           &     6.99 &    1.37 &   10.20 &     50.9 &    14.2 &    18.6 &    14.6 &  -108.0 &     3.5 &   H & YH \\
NGC 5466     &     5.17 &    3.01 &   15.25 &    256.0 &    83.5 &   -51.9 &    83.4 &   199.4 &    24.4 &   H & YH \\
NGC 5897     &    -7.66 &    0.24 &   16.65 &     49.7 &    50.8 &    82.7 &    44.7 &   119.5 &    44.8 &   H & OH \\
M5           &     3.38 &    0.35 &    5.47 &   -324.1 &    34.8 &   113.4 &    30.5 &  -214.1 &    31.7 &   H & OH \\
M80          &    -0.85 &   -1.20 &    3.33 &     92.5 &    31.5 &    86.6 &    35.0 &   -99.5 &    29.9 &   H & OH \\
M4           &     6.41 &   -0.33 &    0.61 &    -51.2 &     3.0 &   -19.7 &    21.9 &   -16.0 &     5.8 &   H & OH \\
M13          &     5.50 &    4.99 &    5.04 &    293.6 &    35.5 &   -89.9 &    35.4 &  -111.8 &    20.3 &   H & OH \\
M12          &     4.27 &    1.19 &    2.17 &    -28.2 &    12.6 &   132.7 &    19.4 &  -132.9 &    17.1 &  TK & OH \\
M10          &     4.59 &    1.06 &    1.72 &    -65.7 &     8.8 &   126.5 &    19.3 &    88.6 &    15.5 &  TK & OH \\
NGC 6287     &    -0.63 &    0.02 &    1.78 &  \nodata & \nodata & \nodata & \nodata & \nodata & \nodata &   B & OH \\
NGC 6293     &    -0.21 &   -0.36 &    1.20 &  \nodata & \nodata & \nodata & \nodata & \nodata & \nodata &   B & OH \\
M92          &     6.02 &    6.25 &    4.69 &     82.4 &    14.4 &   -16.6 &    14.4 &    79.4 &    20.9 &   H & OH \\
NGC 6342     &     0.05 &    0.72 &    1.45 &  \nodata & \nodata & \nodata & \nodata & \nodata & \nodata &   B & BD \\
NGC 6352     &     3.14 &   -1.80 &   -0.71 &  \nodata & \nodata & \nodata & \nodata & \nodata & \nodata &  TN & BD \\
NGC 6362     &     2.53 &   -4.10 &   -2.29 &    -43.0 &    20.2 &   119.4 &    15.5 &    41.4 &    17.1 &  TK & OH \\
NGC 6397     &     6.41 &   -0.84 &   -0.48 &     23.1 &     6.7 &   124.6 &    12.2 &  -108.8 &    11.3 &  TK & OH \\
NGC 6528     &     0.62 &    0.16 &   -0.57 &    189.7 &     4.2 &    57.9 &    10.8 &     4.3 &    10.4 &   B & BD \\
NGC 6541     &     1.75 &   -1.25 &   -1.35 &  \nodata & \nodata & \nodata & \nodata & \nodata & \nodata &   B & OH \\
NGC 6553     &     2.53 &    0.55 &   -0.32 &     55.4 &     4.6 &   218.6 &    16.4 &    13.7 &     2.2 &  TN & OH \\
M22          &     5.37 &    0.54 &   -0.42 &    172.9 &     5.5 &   176.7 &    21.7 &  -122.9 &    24.6 &  TK & OH \\
M54          &   -17.37 &    2.54 &   -6.52 &  \nodata & \nodata & \nodata & \nodata & \nodata & \nodata &   H & SG \\
NGC 6752     &     5.19 &   -1.44 &   -1.73 &    -24.7 &     5.6 &   193.9 &     9.1 &    21.9 &     7.2 &  TK & OH \\
M55          &     3.69 &    0.74 &   -2.09 &   -184.4 &     9.6 &    43.7 &    27.7 &  -108.5 &    15.3 &   H & OH \\
M71          &     6.31 &    3.33 &   -0.32 &      3.1 &    14.1 &   178.7 &    11.4 &    -2.6 &    14.8 &  TN & BD \\
NGC 7006     &    -8.80 &   35.11 &  -13.79 &   -179.6 &    44.4 &   167.1 &    71.3 &   147.8 &    68.5 &   H & YH \\
M15          &     4.63 &    8.33 &   -4.73 &   -142.5 &    23.8 &   167.8 &    34.2 &   -96.0 &    31.1 &   H & YH \\
M30          &     3.63 &    2.50 &   -5.83 &    -15.6 &    25.0 &  -120.5 &    27.2 &    50.4 &     2.0 &   H & OH \\
NGC 7492     &     1.63 &    9.25 &  -23.09 &  \nodata & \nodata & \nodata & \nodata & \nodata & \nodata &   H & OH \\ 
Liller 1     &    -1.06 &   -0.86 &   -0.03 &  \nodata & \nodata & \nodata & \nodata & \nodata & \nodata &   B & BD \\
Palomar 5    &    -7.66 &    0.24 &   16.65 &    -78.4 &    14.2 &   110.3 &    38.9 &    14.3 &    13.9 &   H & YH \\
Palomar 6    &     2.61 &    0.22 &    0.18 &  \nodata & \nodata & \nodata & \nodata & \nodata & \nodata &   B & YH \\
Palomar 12   &    -2.58 &    6.53 &  -14.12 &      1.8 &    39.7 &   243.3 &    31.8 &   -20.9 &    18.3 &   H & SG \\
Ruprecht 106 &    -2.16 &  -17.82 &    4.29 &  \nodata & \nodata & \nodata & \nodata & \nodata & \nodata &   H & YH \\
Terzan 4     &    -0.58 &   -0.63 &    0.21 &  \nodata & \nodata & \nodata & \nodata & \nodata & \nodata &   B & OH \\
Terzan 5     &    -1.77 &    0.69 &    0.30 &  \nodata & \nodata & \nodata & \nodata & \nodata & \nodata &   B & BD \\
Terzan 7     &   -13.25 &    1.29 &   -7.96 &  \nodata & \nodata & \nodata & \nodata & \nodata & \nodata &   H & SG \\
\enddata
\tablecomments{Class1 is Galactic component we found from our calculations as outlined 
in \S3.  Class2 are the Galactic components taken from Mackey \& Gilmore (2004). The 
abbreviations are: B$=$bulge, TN$=$thin disk, TK$=$thick disk, H$=$halo, YH$=$young halo, 
OH$=$old halo, BD$=$bulge/disk, SG$=$Sagittarius cluster.}
\end{deluxetable}


\begin{deluxetable}{lcc}
\tablewidth{0pc}
\tablecaption{Possible Kinematics for Ruprecht~106 \label{tbl-4}}
\tablenum{4} 
\tablehead{
\colhead{Proper Motion} & \colhead{U,V,W} & \colhead{$\Pi$, $\Theta$}  \\[.2ex]
\colhead{RA ("), DEC (")} & \colhead{\kms} & \colhead{\kms} 
          }
\startdata
\phs0,\phs0 & $+118.4$, $+194.4$, $-46.0$ & $-207.2$, $+95.6$ \\
\phs1,\phs1 & $-58.1$, $+336.6$, $+100.7$ & $-326.4$, $-90.9$ \\
$-$1,$-$1   & $+84.9$, $+201.1$, $-104.2$ & $-211.3$, $+59.6$ \\
\phs1,$-$1  & $-85.7$, $+306.9$, $-96.1$ & $-294.8$, $-116.6$ \\
$-$1,\phs1  & $+112.6$, $+230.8$, $+92.7$ & $-242.9$, $+85.3$ \\
\enddata
\end{deluxetable}

\end{document}